\documentclass[preprint,prd,nofootinbib,tightenlines,amsmath]{revtex4}
\usepackage{enumerate}
\usepackage{multirow}

\usepackage{epsfig}
\usepackage{graphics,color}      
\usepackage{verbatim}
\usepackage{amsmath}    
\catcode`\@=11
\def\sla@#1#2#3#4#5{{%
 \setbox\z@\hbox{$\m@th#4#5$}%
 \setbox\tw@\hbox{$\m@th#4#1$}%
 \dimen4\wd\ifdim\wd\z@<\wd\tw@\tw@\else\z@\fi
 \dimen@\ht\tw@
 \advance\dimen@-\dp\tw@ \advance\dimen@-\ht\z@
 \advance\dimen@\dp\z@
 \divide\dimen@\tw@ \advance\dimen@-#3\ht\tw@
 \advance\dimen@-#3\dp\tw@ \dimen@ii#2\wd\z@
 \raise-\dimen@\hbox to\dimen4{%
 \hss\kern\dimen@ii\box\tw@\kern-\dimen@ii\hss}%
 \llap{\hbox to\dimen4{\hss\box\z@\hss}}}}

\def\slashed#1{%
 \expandafter\ifx\csname sla@\string#1\endcsname\relax
{\mathpalette{\sla@/00}{#1}}
\fi}
\def\declareslashed#1#2#3#4#5{%
 \expandafter\def\csname sla@\string#5\endcsname{%
#1{\mathpalette{\sla@{#2}{#3}{#4}}{#5}}}}
 \catcode`\@=12
\declareslashed{}{/}{.08}{0}{D}
 \declareslashed{}{/}{.1}{0}{A}
 \declareslashed{}{/}{0}{-.05}{k}
 \declareslashed{}{/}{.1}{0}{\partial}
 \declareslashed{}{\not}{-.6}{0}{f}

\def\lsim{\mathrel {\vcenter {\baselineskip 0pt \kern 0pt
    \hbox{$<$} \kern 0pt \hbox{$\sim$} }}}
\def\gsim{\mathrel {\vcenter {\baselineskip 0pt \kern 0pt
    \hbox{$>$} \kern 0pt \hbox{$\sim$} }}}

\def\zp{Z^\prime}
\newcommand{\bea}{\begin{eqnarray}}
\newcommand{\eea}{\end{eqnarray}}

\begin{document}

\baselineskip=15pt
\preprint{}

\title{Same sign top-pairs in a non-universal $\zp$ model at the LHC}

\author{Sudhir Kumar Gupta}

\email{skgupta@iastate.edu}

\affiliation{Dept of Physics \& Astronomy, Iowa State University, Ames, IA 50011.}

\date{\today}

\vskip 1cm
\begin{abstract}

We analyse same sign dilepton signatures in a non-universal flavor changing 
$\zp$ model. These arise due to $tt$ (or $\bar{t}\bar{t}$) production processes due 
to the semi-leptonic decays of (anti)tops. We also discuss top reconstruction 
and spin measurement using the variable {\em $M_{T_2}$} and {\em 
$M_{T_2}$-Assisted On-Shell (MAOS) Momentum} techniques and will also provide 
a comparison with the on-shell mass relation method. Sensitivities to the 
flavor-changing top coupling has also been estimated for different LHC energies and 
projected LHC luminosities corresponding to them.

\end{abstract}

\pacs{PACS numbers: }

\maketitle

\section{Introduction}

Presence of extra $\zp$ bosons is dictated by a wide range of extensions 
to the Standard Model (SM). These arise due to presence of additional 
abelian gauge symmetries $U(1)$, as part of extended SM gauge groups 
$G_{SM} \times {U(1)}^N$; N=1,2,...~\cite{review}. Phenomenology of such 
models is interesting as these couples to the SM fermions with flavor-diagonal as well as off-diagonal couplings. A $\zp$.  which couples to SM fermions with flavor violating couplings is even 
interesting as it, besides contributing to large top-quark 
forward-backward asymmetry $A^t_{FB} = .193\pm .069$, as measured at the 
Tevatron, tree-level quark sector Flavor-changing neutral currents 
(FCNCs)~\cite{fcnczp} which were suppressed in the SM, also give rise to 
interesting collider signatures such as same sign top pairs, associated production of $\zp$ with a 
top or antitop\cite{Gupta:2010wt}.   

In traditional $\zp$ model, flavor violating coupling to the quarks are tiny so the process $tt$ is irrelevant in those cases. However it has been recently argued in the Refs.~\cite{Jung:2009jz}, \cite{BarShalom:2007pw} that at least some of these flavor off-diagonal couplings can be comparable to $V_{tb}$ in models where a right chiral $\zp$ couples in a non-trivial way to the 
up-quarks.

The aim of this paper is to study in detail the like sign signatures in the context of the Large Hadron Collider (LHC) which arise via the same sign top pair production. This is interesting as it will serve as a direct probe to the nature of $\zp$ and its coupling to the quarks. 

Organisation of the article is as follows: In the next Section we will 
briefly discuss about the model and its experimental constraints. In 
Section 3, we will discuss the top and anti-top pair production 
cross-section for a wide range of $\zp$ mass. We will work with same 
sign dilepton signature at the LHC, top reconstruction, and spin 
measurment of $\zp$ in Section 4. We will also discuss LHC sensitivities 
to the model in the same Section. Finally we will summarise our findings 
in Section 5.

\section{The Model}

As has been discussed in the previous section, in our model, the new vector boson $\zp$ couples with the up quarks via the right handed coupling with the following interaction terms

\begin{equation} 
{\mathcal L} \ni G_{ij} \zp_{\mu} 
\bar{Q}_{iL} \gamma^{\mu} P_{R} u_{jR} +h.c. \label{eq:zpcoup} 
\end{equation} 

where $G_{ij}$ is a $3\times 3$ matrix of the form

\[
\left(
\begin{array}{ccc}
0  & 0  & \eta_{13}  \\
 0 & 0  & \eta_{23}  \\
\eta_{31}  & \eta_{32}  & 0   
\end{array}
\right)
\]

It has been been pointed out by the authors of 
Ref.~\cite{BarShalom:2007pw}, that the couplings $\eta_{31}$, 
$\eta_{32}$ can be $\sim V_{tb} >> \eta_{33} \sim V_{td,s} $ which is 
consistent with the low energy flavor data such as meson mixing and 
$K-decays$ \cite{fcncconstraints}. It has also been discussed in the 
same article that the model will be free from any such constraints 
provided only one flavor violating coupling is non-zero. In our study we 
will not restrict ourselves with the aforementioned coupling to be $\sim 
{\mathcal O} (1)$, but will rather study a whole range with $\eta_{31} 
\ne 0$. Thus, the relevant interaction term will take the following form

\begin{equation}
{\mathcal L} \ni g_{_X} \zp_{\mu} \bar{u} \gamma^{\mu} P_{R} t +h.c.
\end{equation}

with $g_{_X} \in (0,1]$.

\section{(Same-sign) top pair production}

\begin{figure}
\centerline{
\includegraphics[angle=0, width=1\textwidth]{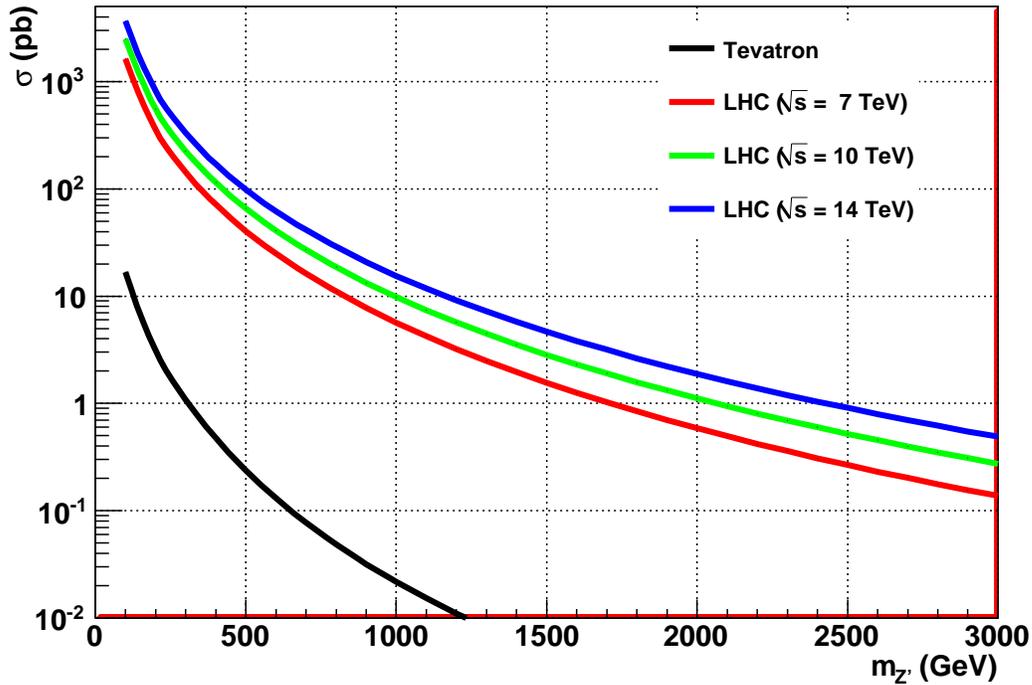}
}
\caption{\small\sf Tevatron and LHC cross-sections for same-sign top 
pair productions at the Tevatron and LHC with $\sqrt{S}$ as $1.98$ 
(Black), $7$ (Red), $10$ (Green) and $14$ TeV (Blue) respectively. $g_X 
=1$ is assumed here.}
\label{fig:sigma}
\end{figure}

If the flavor violating coupling $u-t-\zp$ is sufficiently 
large, we expect to observe plenty of same sign top pairs at the LHC depending upon its coupling. With the setup we have, the only responsible subprocesses for the 
same sign (anti)top pairs are $\bar{u}\bar{u}\to \bar{t} \bar{t}$ and 
$uu\to tt$. These occurs by the t- (and u-) channel exchange of the 
$\zp$. 

We present total cross-section for the processes $tt+\bar{t}\bar{t}$ at 
the LHC and the Tevatron for a wide range of $\zp$ mass between $0.1-3$ 
TeV in Fig.~\ref{fig:sigma}. We use {\tt CTEQ6L1} to estimate the parton densities
 in our cross-section calculation. The two QCD scales, i.e. the renormalization scale, 
$\mu_R$, and the factorization scale $\mu_F$ are fixed at 

\begin{equation}
\mu_R = \sqrt{\hat s} = \mu_F.
\end{equation}

We do not use the K-factors in our cross-section calculation. If these 
are similar to as given in the Ref.~\cite{Kidonakis:2003sc}, the rates 
are expected to go up by about $\sim 20\%$ at {\em NNLO-NLL}.

\begin{figure}
\centerline{   
\includegraphics[angle=-90, width=1\textwidth]{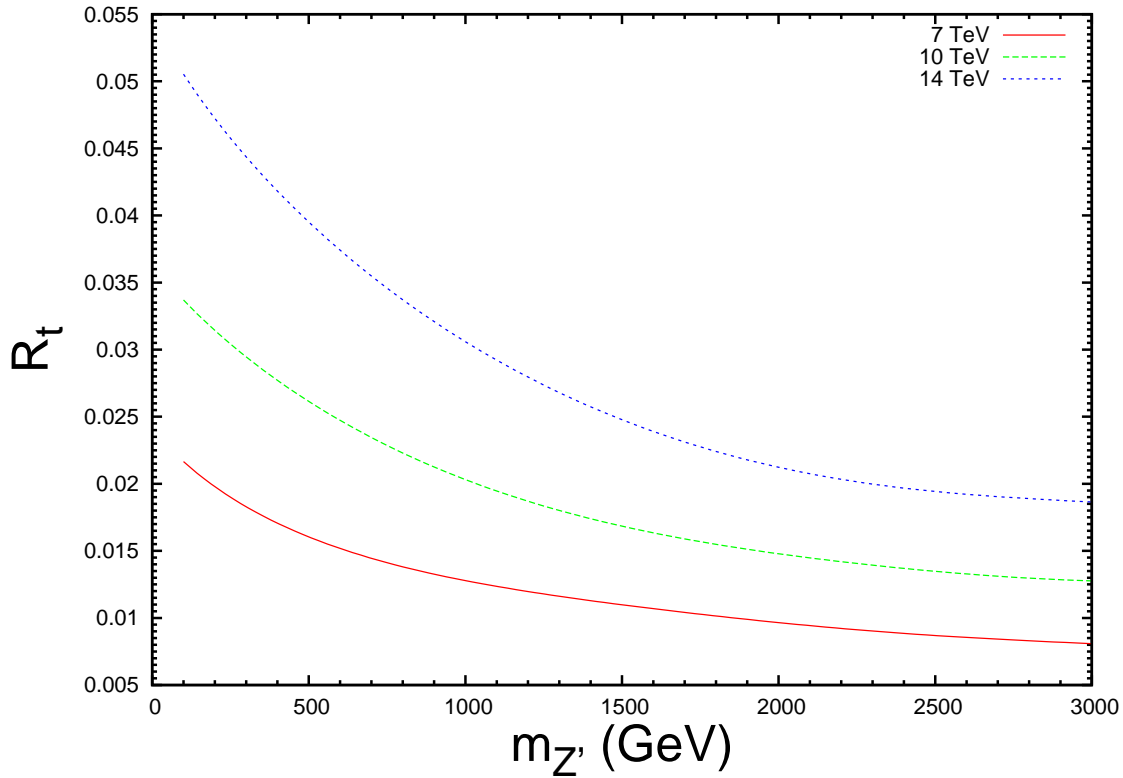}
}

\caption{\small\sf Ratio $R_t = \sigma_{\bar {t}\bar {t}}/ \sigma_{tt}$ 
vs $m_{Z'}$ at the LHC for $\sqrt{S} =$ 7, 10 and 14 TeV. Color 
convention is the same as in Figure~\ref{fig:sigma}.}
\label{fig:ratio}
\end{figure}

It is worth to note here that, at the Tevatron the contribution to 
the total cross-section due to $tt$ and $\bar{t}\bar{t}$ are the same, 
i.e.  $R_t = \sigma_{\bar {t}\bar {t}}/ \sigma_{tt} =1$. This is 
because, at the Tevatron, in both production processes one parton is 
always a valence up-quark while the other is a sea up-quarks. However at 
the LHC the situation is quite different, i.e. in one case (tt) both the 
partons are either valence quarks while in other $\bar{t}\bar{t}$ these are sea 
quarks. Clearly, we expect, $R_t$ to be $ < 1$ at the LHC. (See 
Figure~\ref{fig:ratio})

A couple of interesting remarks about these ratios: (a) $R_t$'s are 
independent of the coupling constant, and, also (b) independent of 
higher order QCD and electroweak correction as these corrections will be 
exactly same for both $\bar{u}\bar{u}\to \bar{t} \bar{t}$ and $uu\to 
tt$, thus cancel between the denominator and the numerator. Another 
interesting feature about these ratios is that these will remain intact 
for a similar detection mode due to cancelation of branching ratios 
between tops and antitops.

Clearly such ratios can serve as an important tool to probe 
$m_{z^\prime}$ in addition to other kinematic variables.

\section{Same sign dileptons at the LHC}

With their cross-sections given as in Figure~\ref{fig:sigma}, {\em a 
priori}, we will have enough events well above the Tevatron reach of 
same sign top pairs with $g_X \sim 0.01$ for $m_{\zp} = 300$ GeV and 
$g_{_X} \sim 0.2$ for $m_{\zp} = 1.2$ TeV respectively, in case both 
tops are fully reconstructed in all their decay modes.

Semileptonic decays of the produced tops give rise to a very striking form of 
LHC signature in the form of a pair of same sign leptons accompanied by 
a pair of b-jets and some missing energy, $\slashed{E}_T$ due to missing 
neutrinos from the decay of each top. These same sign dileptons are 
expected to serve as a remarkable probe to the new physics models where 
the decay chain of pair produced new resonances can lead to 
dileptons~\cite{Dreiner:1993ba}. In our study also, this is a unique 
signature to the same sign (anti-)top pairs with almost negligible SM 
background.

For our analysis, we generated top pair events using {\tt MadGraph} 
\cite{Stelzer:1994ta,Alwall:2007st,Alwall:2008pm}.  we produced MadGraph 
model files that incorporates the new particle $Z^\prime$ and the FCNC 
couplings of Eq.~\ref{eq:zpcoup} into {\tt MadGraph} \footnote{These 
modifications are available upon request.}.

In order to study the process, $pp\to l^{^\pm} l^{^\pm} + b b + \slashed 
E_T$, we choose $m_{z^\prime} = .5, 1$ and 1.5 TeV.  The coupling 
$g_{_X}$ is fixed at unity so that for a given $g_{_X}$, the event rates 
can be easily obtained simply by multiplying the factor ${g_{_X}}^4$.  
We present our results for the present LHC centre-of-mass (CM) energy, 
$\sqrt{S} =$ 7 TeV as well as for 10 TeV and 14 TeV.

The event analysis is performed with {\tt 
PYTHIA}~\cite{Sjostrand:2006za} at the parton level, turning off 
initial- and final-state radiation. To select our same sign dilepton 
(SSD) states, we impose the following minimal acceptance cuts on our 
event samples:

\begin{itemize}

\item Both lepton should have $p_{T_l} > 25$ GeV and 
$\left|\eta_{_l}\right| \leq 2.7$, to ensure that they lie within the 
coverage of the detector.

\item b-jets should have $p_{T_b} > 25$ GeV and $\left|\eta_{_b}\right| 
\leq 2.5$

\item Spatial resolution between {\em lepton - lepton}, {\em lepton - 
b-jet}, and, {\em b-jet - b-jet} should be $\Delta{R}_{ll} \geq 0.4$, 
$\Delta{R}_{lb} \geq 0.4$, $\Delta{R}_{bb} \geq 0.4$ respectively, 
(where $\Delta{R}_{ij} = \sqrt{{\Delta{\eta}_{_{ij}}}^2 
+{\Delta{\phi}_{_{ij}}}^2 }$, $\Delta{\eta}_{_{ij}} = \eta_{_i} - 
\eta_{_j}$, $\Delta{\phi}_{_{ij}} = \phi_{_i} - \phi_{_j}$), such that 
the leptons are well separated in space.

\item A missing transverse energy cut, $\slashed E_T > 30~{\rm GeV}$ to 
ensure that leptons are due to $W$ decay.

\end{itemize}

\begin{figure}
\centerline{
\includegraphics[angle=0, width=.4\textwidth]{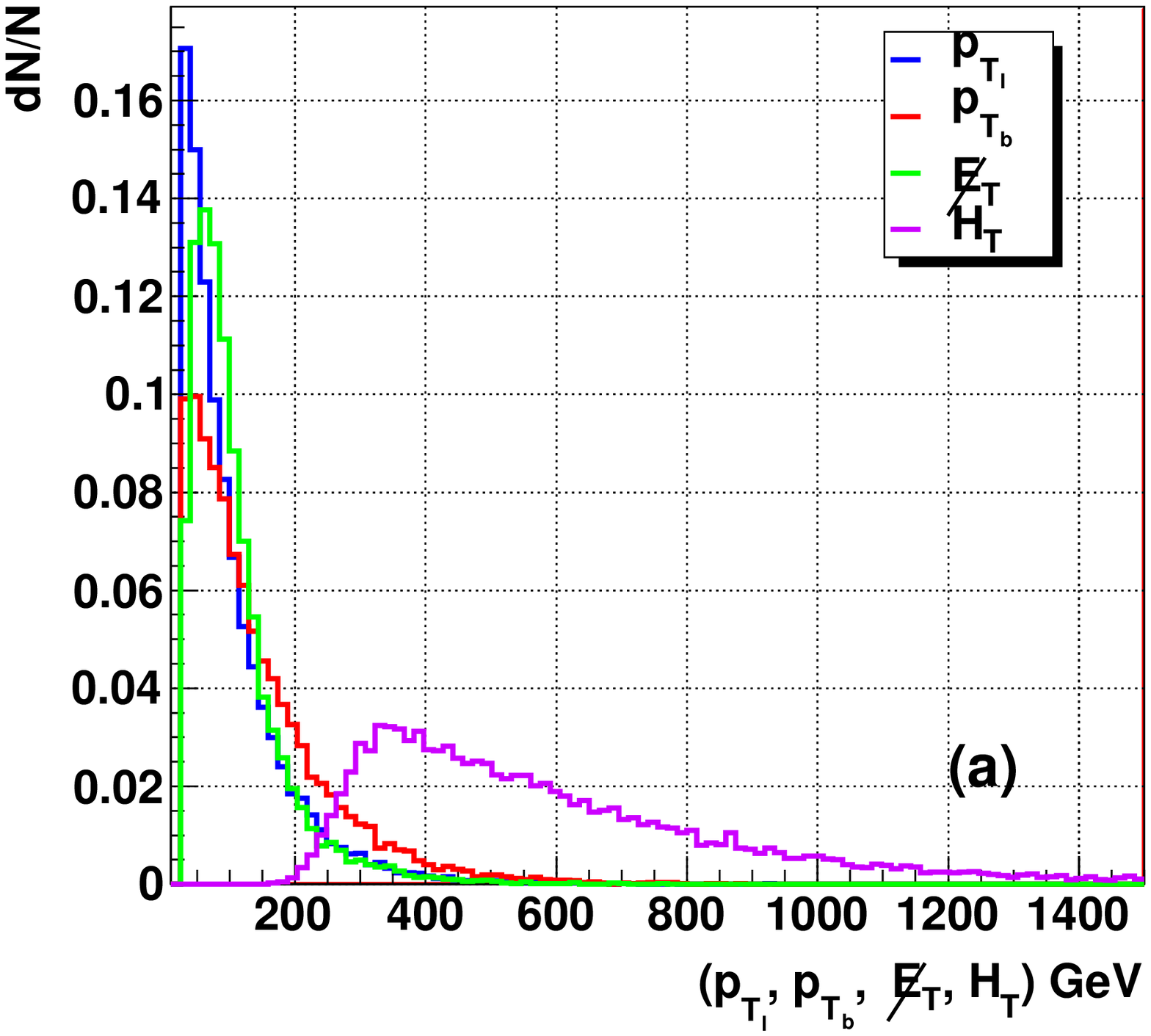}
\includegraphics[angle=0, width=.4\textwidth]{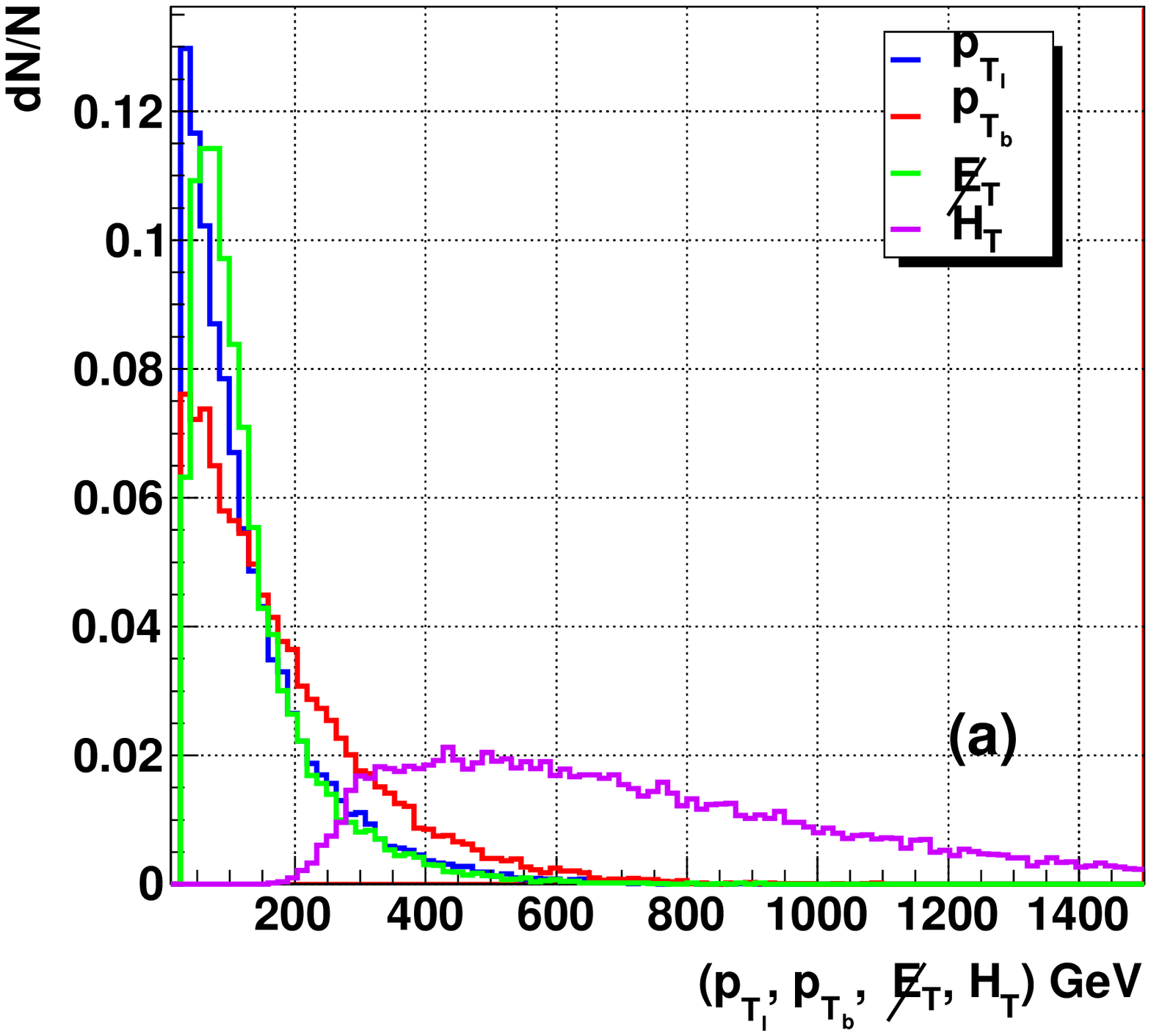}
\includegraphics[angle=0, width=.4\textwidth]{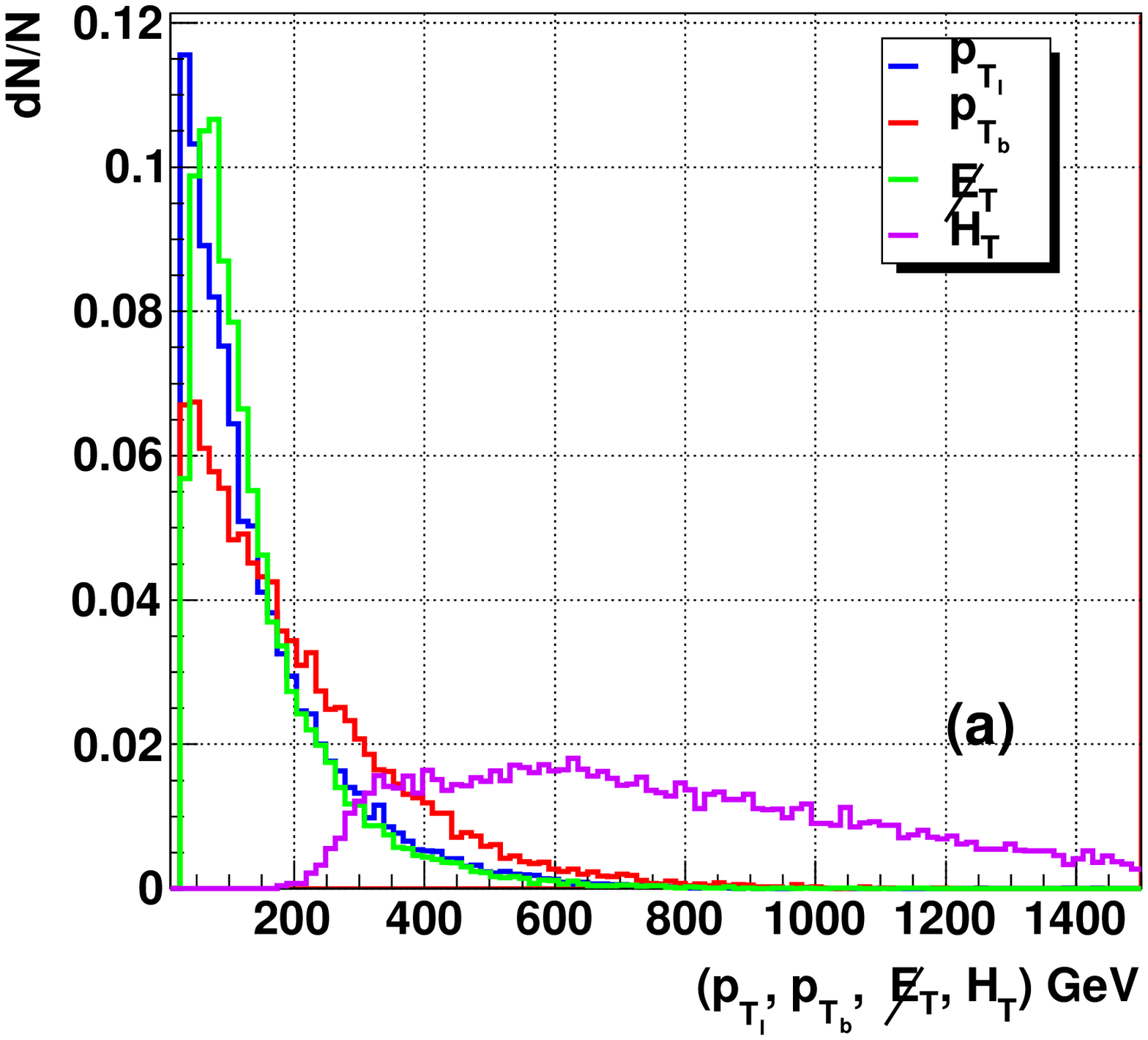}   
}

\caption{\small\sf Differential distributions for lepton and b-jet-$p_T$ 
($p_{T_{l,b}}$), missing engery $\slashed{E}_T$, and scalar-$p_T$ 
($H_T$). $m_{Z'} = 0.5, 1$ and $1.5$ TeV in Figures (a), (b) and (c) 
respectively. $\sqrt{S} = 7$ TeV is assumed here.}
\label{fig:lhc7}
\end{figure}

\begin{figure}
\centerline{
\includegraphics[angle=0, width=.4\textwidth]{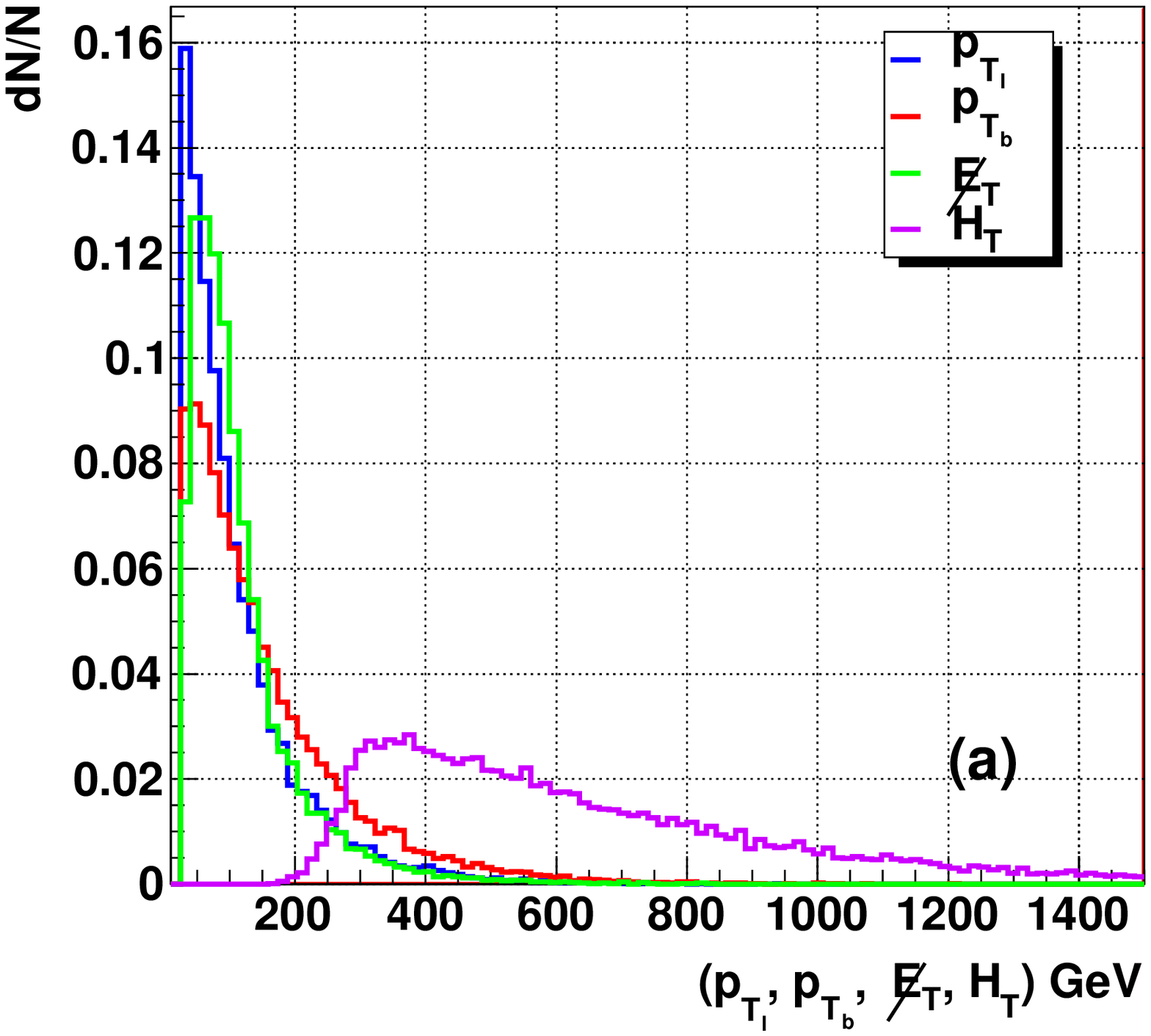}
\includegraphics[angle=0, width=.4\textwidth]{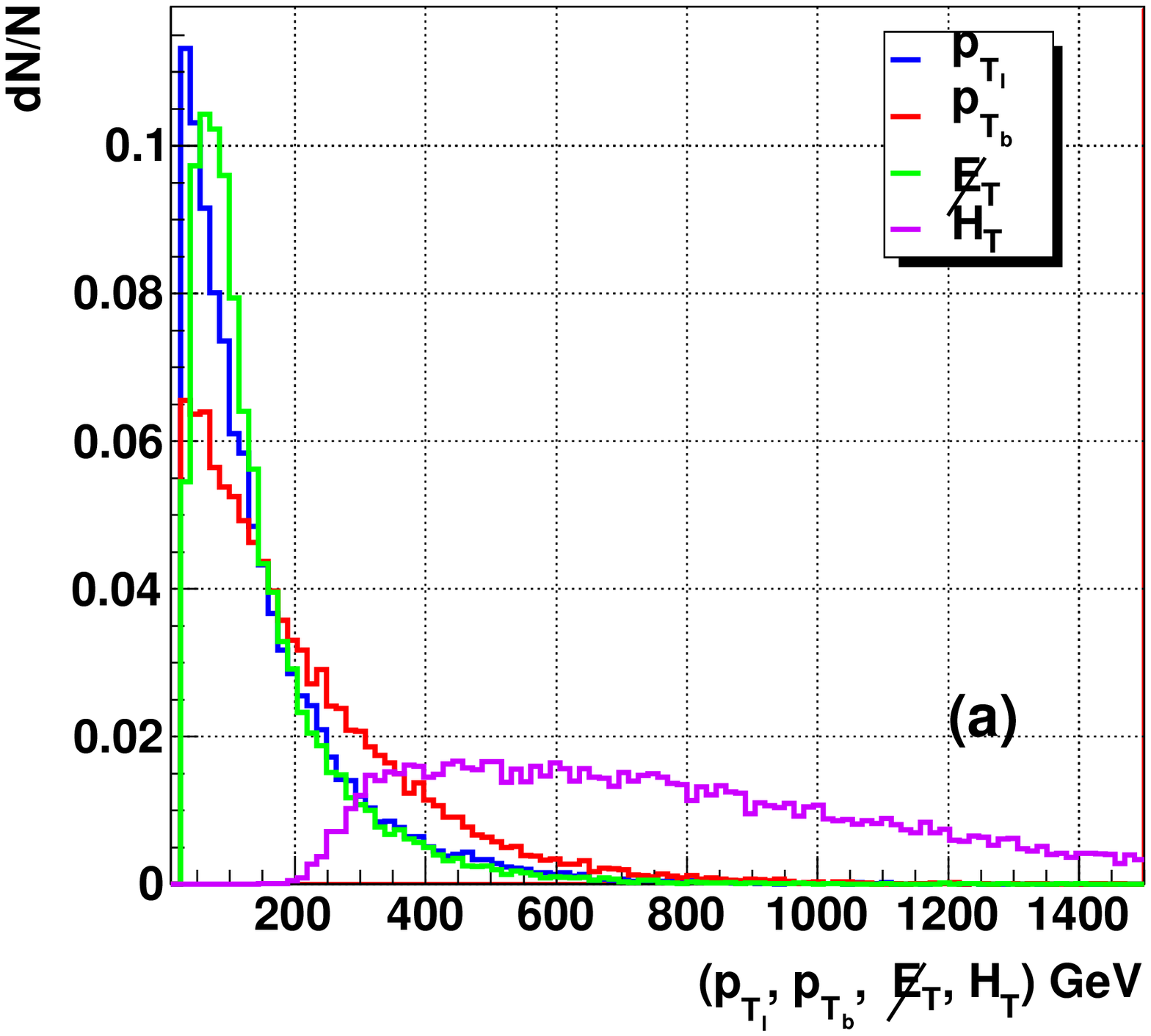}
\includegraphics[angle=0, width=.4\textwidth]{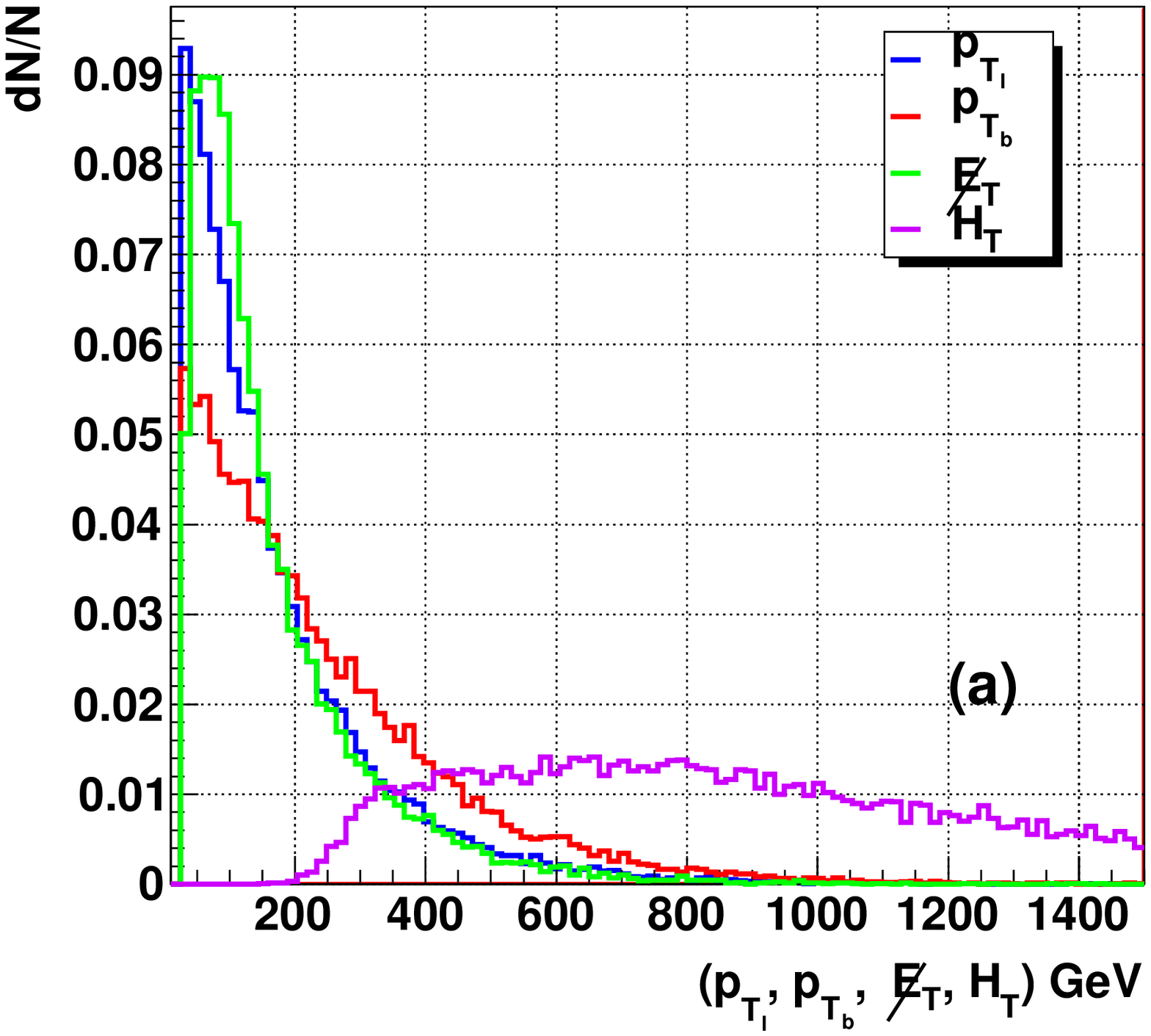}
}
\caption{\small\sf Differential distributions for lepton and b-jet-$p_T$ 
($p_{T_{l,b}}$), missing engery $\slashed{E}_T$, and scalar-$p_T$ 
($H_T$). $m_{Z'} = 0.5, 1$ and $1.5$ TeV in Figures (a), (b) and (c) 
respectively. $\sqrt{S} = 10$ TeV is assumed here.}
\label{fig:lhc10}
\end{figure}

\begin{figure}
\centerline{
\includegraphics[angle=0, width=.4\textwidth]{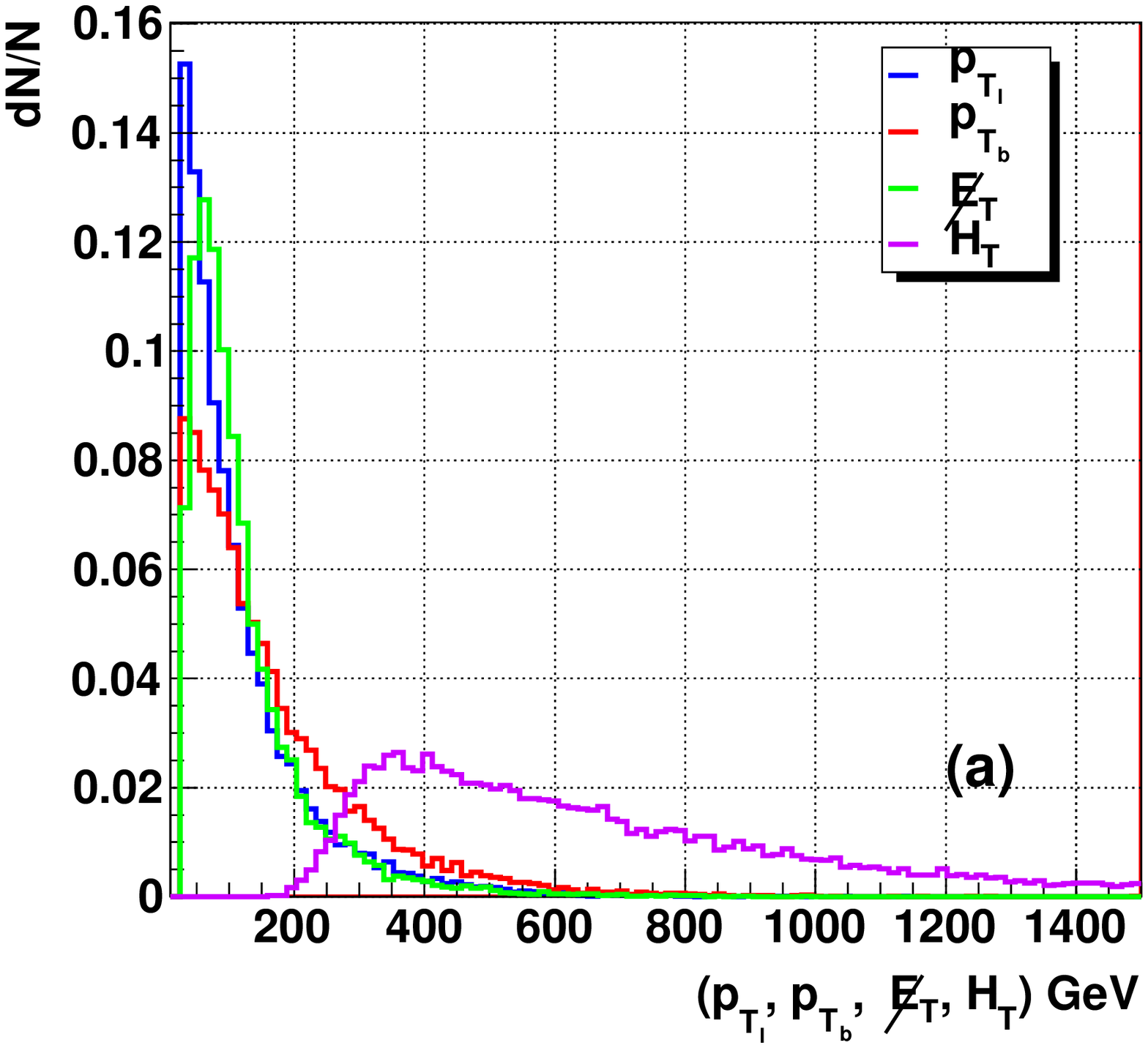}
\includegraphics[angle=0, width=.4\textwidth]{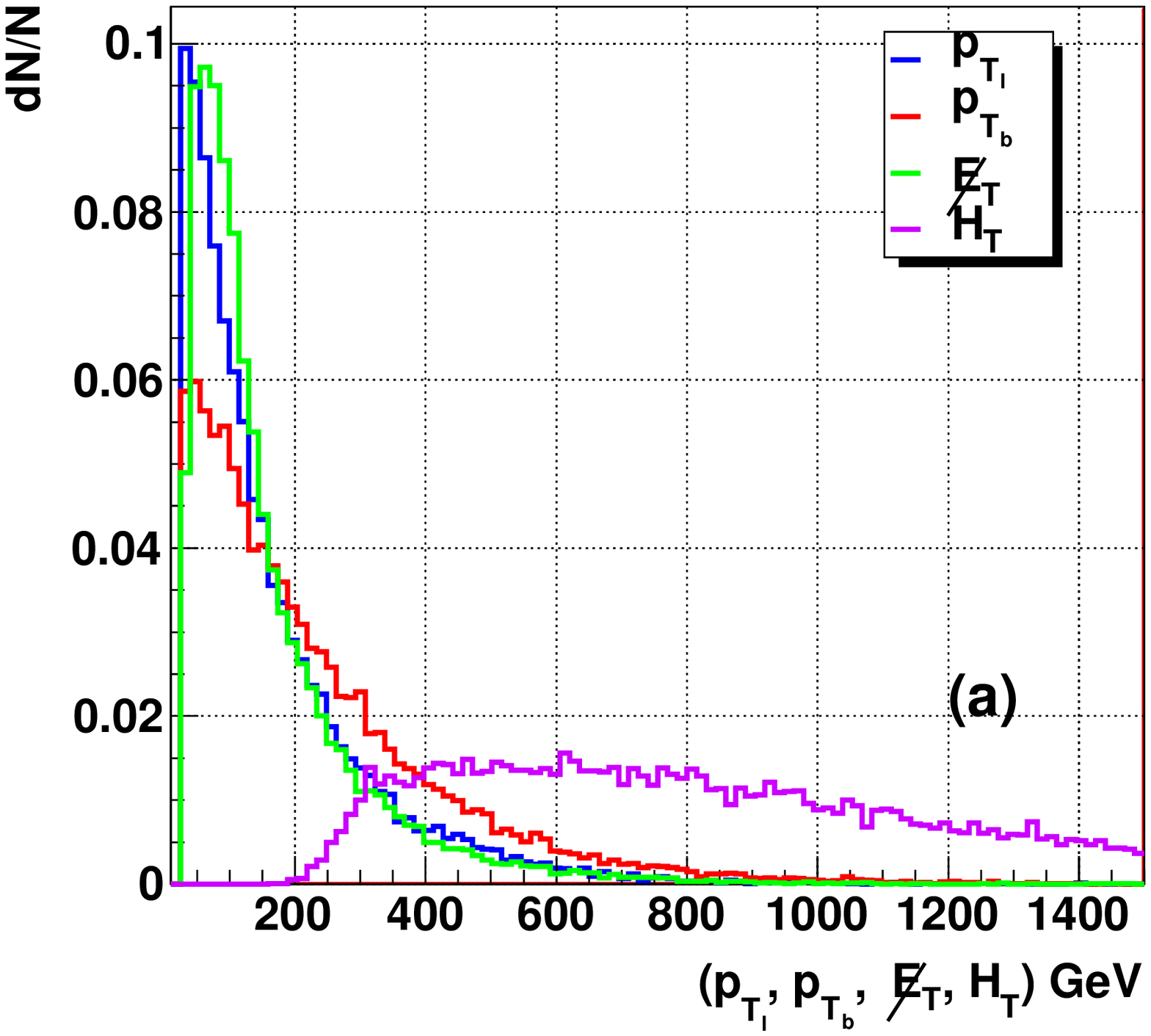}
\includegraphics[angle=0, width=.4\textwidth]{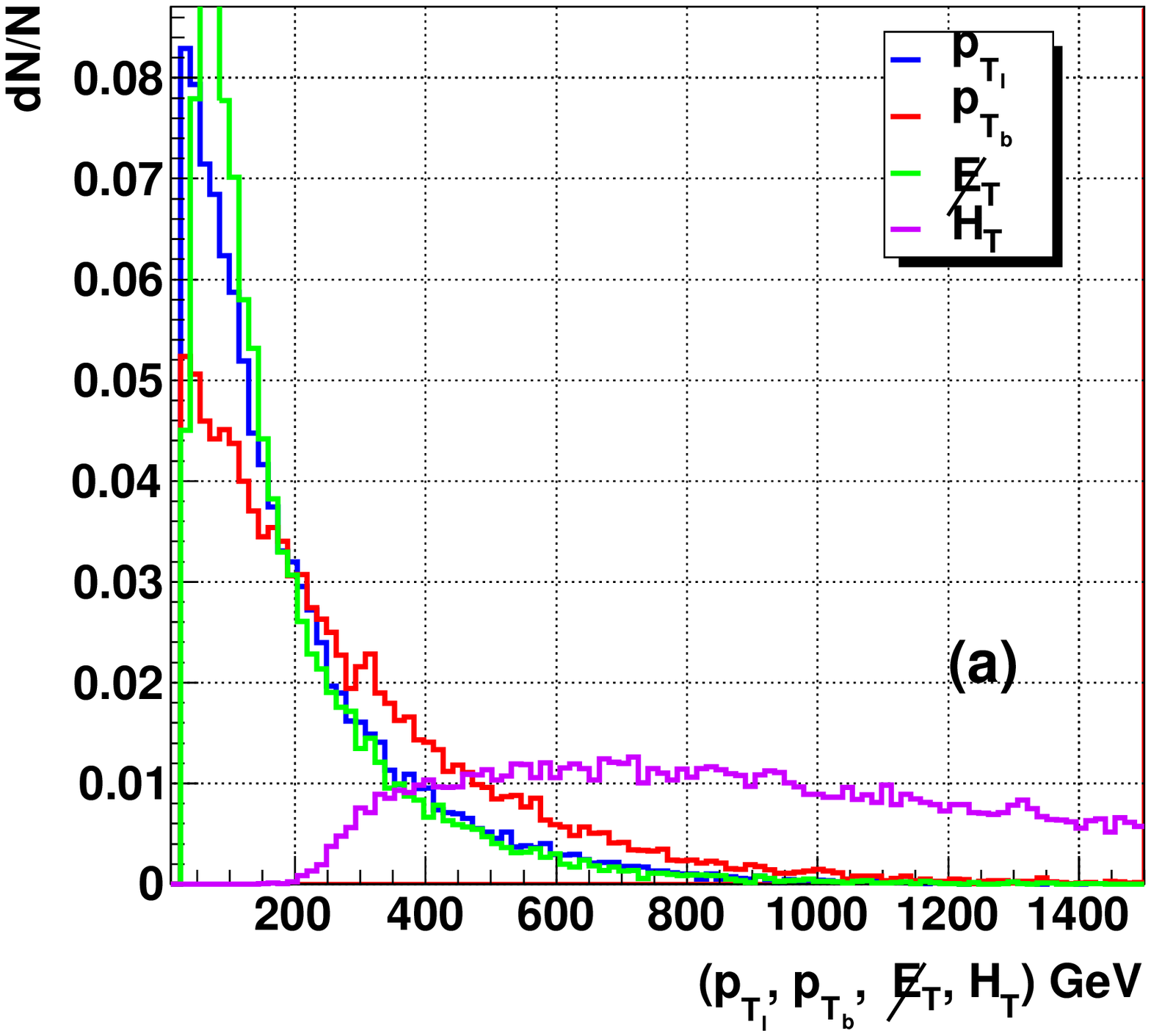}
}
\caption{\small\sf Differential distributions for lepton and b-jet-$p_T$ 
($p_{T_{l,b}}$), missing engery $\slashed{E}_T$, and scalar-$p_T$ 
($H_T$). $m_{Z'} = 0.5, 1$ and $1.5$ TeV in Figures (a), (b) and (c) 
respectively. $\sqrt{S} = 14$ TeV is assumed here.}
\label{fig:lhc14}
\end{figure}

In our event analysis we also allow leptonic decays of $\tau^{\pm}$ 's 
into $e^{\pm}$ or $\mu^{\pm}$. Though the lepton arising from the $\tau$ 
decays are relatively softer, yet they can contribute by $\sim3\%$ in 
the total event rates. Finally, we also used the b-tagging efficiency 
$\sim 58$ percent~as expected in the ATLAS and CMS 
experiments~\cite{Ball:2007zza}.

We present kinematical distributions for lepton and b-jet transverse momentum, 
missing energy and the scalar sum of $p_T$'s of all the visible final 
state particles and the missing transverse energy, i.e.

\begin{equation} 
H_T = p_{_{T_{vis}}} + \slashed E_T = \sum_{l,b} p_T + 
\left|\slashed{\mathbf{p}}_T \right| = \sum_{l,b} p_T + 
\left|-\sum_{l,b} \mathbf{p}_T \right| \end{equation} for different LHC 
energies in Figs.~\ref{fig:lhc7}-\ref{fig:lhc14} at the LHC for $\sqrt{S} =$ 7, 10, and, 14 TeV for three different values of $\zp$ mass in each case as 0.5, 1 and 1.5 TeV. We normalised our distribution with the total number of events in each case.  Though finally it is irrelevant what value of $g_{_X}$ we choose in these normalised distributions as the factor $g^2_{_X}$ will cancel between the numerator and the denominator, we use $g_{_X} = 1$  in our simulation. 

\begin{table}[b]
\begin{tabular}{|l|l|l|l|}
\hline\hline
{\bf $\sqrt{S}$ (TeV), $\int{\cal L} dt$ (fb$^{-1}$)} & {\bf $m_{Z'} = 0.5$ TeV} & {\bf $m_{Z'} = 1$ TeV} & {\bf $m_{Z'} = 1.5$ TeV}  \\ \hline\hline	
\multirow{1}{*}{~7, 0.1} & ~~27 (~~0, ~~27) & ~~~4 (~0, ~~~4)& ~~1 (0, ~~1)\\ \hline
\multirow{1}{*}{10, 0.5} & ~221 (~~5, ~216) & ~~40 (~1, ~~39)& ~12 (0, ~12)\\ \hline
\multirow{1}{*}{14, 10~} & 6690 (252, 6438) & 1268 (36, 1232)& 397 (9, 388)\\ \hline\hline
\end{tabular}
\caption{Number of SSD events at the LHC for $m_{Z'} =$ 0.5, 1 and 1.5 
TeV at the LHC for $\sqrt{S}$ as 7, 10 and 14 TeV. $g_X =1$ is assumed 
in this table. Also shown are the number of events with $l^-l^- + 
\bar{b}\bar{b} +\slashed{E}_T$ and $l^+l^+ + {b}{b} +\slashed{E}_T$ 
events respectively, inside the bracktes.}
\label{t:rates}
\end{table}

Keeping in mind about future LHC runs with different amount of data, we 
present SSD event rates in Table~\ref{t:rates} for both the processes 
$l^-l^- + \bar{b}\bar{b} +\slashed{E}_T$ and $l^+l^+ + {b}{b} 
+\slashed{E}_T$ as well as the sum of the two. It is to be noted that 
once including the available K-factor at NNLO-NLL, our predictions for 
the event rate will go up by a factor of $1.2$.

With the events with the aforementioned kinematical distribution, our next 
task is to confirm whether such signatures are really due to the top-pair 
production. Also, once the top are reconstructed the next level question to 
ask is about the nature of the exchanged particle. Keeping this in mind, In 
the remaining part of the section we will deal with issues such as top pair 
reconstruction, which is analogous to reconstructing four-momenta of the 
missing neutrino pair.  Later use the information to reconstruct complete 
subprocess in order to probe the mass and spin information of the exchanged 
$\zp$. Let us begin with the top mass reconstruction in the next subsection.

\subsection{Top Reconstruction}

Since in these final states, the missing transverse energy is mostly composed 
of two invisible neutrinos, it is not obvious to reconstruct tops. Yet, due to 
the fact the produced particles and their decay chains are identical, it is 
still possible to reconstruct them fully up to a finite degree of accuracy 
through the following two methods. These are: (a) Mass relation method (MRM), 
and, (b) $M_{T_2}$-Assisted On-Shell Momentum (MAOS) method. Below we discuss 
them one by one in detail in the present context:

\subsubsection{Mass relation Method}

In this method we use the known on-shell mass relations involving 
four-momenta of various final state particles and make use of the two 
missing transverse momentum relations. Thus, we have

\begin{subequations}
\begin{eqnarray}
p_{\nu_1}^2 &=& 0 \\
p_{\nu_2}^2 &=& 0 \\
{(p_{l_1} + p_{\nu_1})}^2 &=& {m^2_W} \\
{(p_{l_2} + p_{\nu_2})}^2 &=& {m^2_W} \\
{(p_{l_1} + p_{b_1} + p_{\nu_1})}^2 &=& {m^2_t} \\
{(p_{l_2} + p_{b_2} + p_{\nu_2})}^2 &=&{m^2_t}\\
\mathbf{p}_{T_{\nu_1}} + \mathbf{p}_{T_{\nu_2}} &=& \slashed{\mathbf{p}}_T = - \sum_{l,b} \mathbf{p}_T
\end{eqnarray}
\end{subequations}

\subsubsection{$M_{T_2}$-Assisted On-Shell Momentum (MAOS) Technique}

Though, as we will see later that the previous method works fine, yet it has a 
major drawback, i.e. we need to use mass of the top explicitly in the 
aforementioned mass relations. Recently a new method, call 
MT2-method~\cite{mt2} has been found to overcome this problem. This method 
uses the mass-relations in a slightly different way to first define the 
variable $m_{T_2}$ as

\begin{equation}
{M_{T_2}(m_{\mathcal U})} =
\min_{\mathbf{p}_{T}^{(1)}, \mathbf{p}_{T}^{(2)}} \left[ {\max \left\{M_T\left(m_{\mathcal U}; \mathbf{p}_{T}^{(1)}\right),
M_{T}\left(m_{\mathcal U}; \mathbf{p}_{T}^{(2)}\right)\right\}} \right],
\label{eq_min}
\end{equation}

where $M_{T}$, the transverse mass of each parent particle, is defined 
as

\begin{equation}
M_T(m_{\mathcal U}; \mathbf{p}_T^{\mathcal U}) = 
\sqrt{m_{\mathcal V}^2 + m_{\mathcal U}^2 + 2
(E_{T}^{\mathcal V} E_{T}^{\mathcal U} -
\mathbf{p}_{T}^{\mathcal V} \cdot \mathbf{p}_{T}^{\mathcal U})}.\label{eq_mT}
\end{equation}

Here $\mathcal U$ and $\mathcal V$ represent the individual 
undetected~(invisible) and detected~(visible) particles, respectively, 
$\mathbf{p}_T^{(1)}$ and $\mathbf{p}_T^{(2)}$ are transverse momenta of 
two invisible particles and $m_{\mathcal U}$ is the mass of the 
invisible particle. The minimization is performed with the constraint 
$\mathbf{p}_{T}^{(1)}+\mathbf{p}_{T}^{(2)} = \mathbf{\slashed{p}}_{T}$.

One interesting thing about this method is that mass of the top is determined 
before the determinantion of longitudianl momentum of the invisible neutrinos 
which is due to the fact that we are dealing with transverse masses. 

Now, once we obtained transverse momenta of the missing neutrinos through the 
aforementioned way as, $\mathbf{p}_{T}^{\mathcal U_i} = 
\mathbf{p}_{T}^{(i)}$, we can obtain the longitudinal components by 
solving,

\bea 
\label{eq_maos} {p}^{\mathcal U}_{L}=
\frac{1}{(E^{\mathcal V}_{T})^2}\left[ {\mathcal A}~p^{\mathcal V}_{L}\pm
\sqrt{(p_{L}^{\mathcal V})^2+(E_{T}^{\mathcal V})^2}\sqrt{{\mathcal A}^2-(E^{\mathcal V}_{T}E^{\mathcal U}_{T})^2}\right]
\eea 
where

$E^{\mathcal V}_{T}=\sqrt{(p^{\mathcal V})^2+|\mathbf{p}^{\mathcal 
V}_{T}|^2}$, $E^{\mathcal U}_{T}=\sqrt{(p^{\mathcal 
U})^2+|\mathbf{p}^{\mathcal U}_{T}|^2}$, and ${\mathcal 
A}=\frac{1}{2}\left\{m_{\cal P}^2-m_{\mathcal U}^2-(p^{\mathcal 
V})^2\right\} + \mathbf{p}_{T}^{\mathcal V} \cdot 
\mathbf{p}_{T}^{\mathcal U}$. $m_{\mathcal P}$, $m_{\mathcal U}$ are the 
masses of produced particle and the invisible particle respectively.

In our case the top mass and the unknown neutrino momenta are obtained 
by setting, ${\mathcal P} = t$, ${\mathcal U} = \nu$ and ${\mathcal V} = 
b + l$ in the aforementioned equations~\ref{eq_min}-\ref{eq_maos}.

\begin{figure}
\centerline{
\includegraphics[angle=0, width=.4\textwidth]{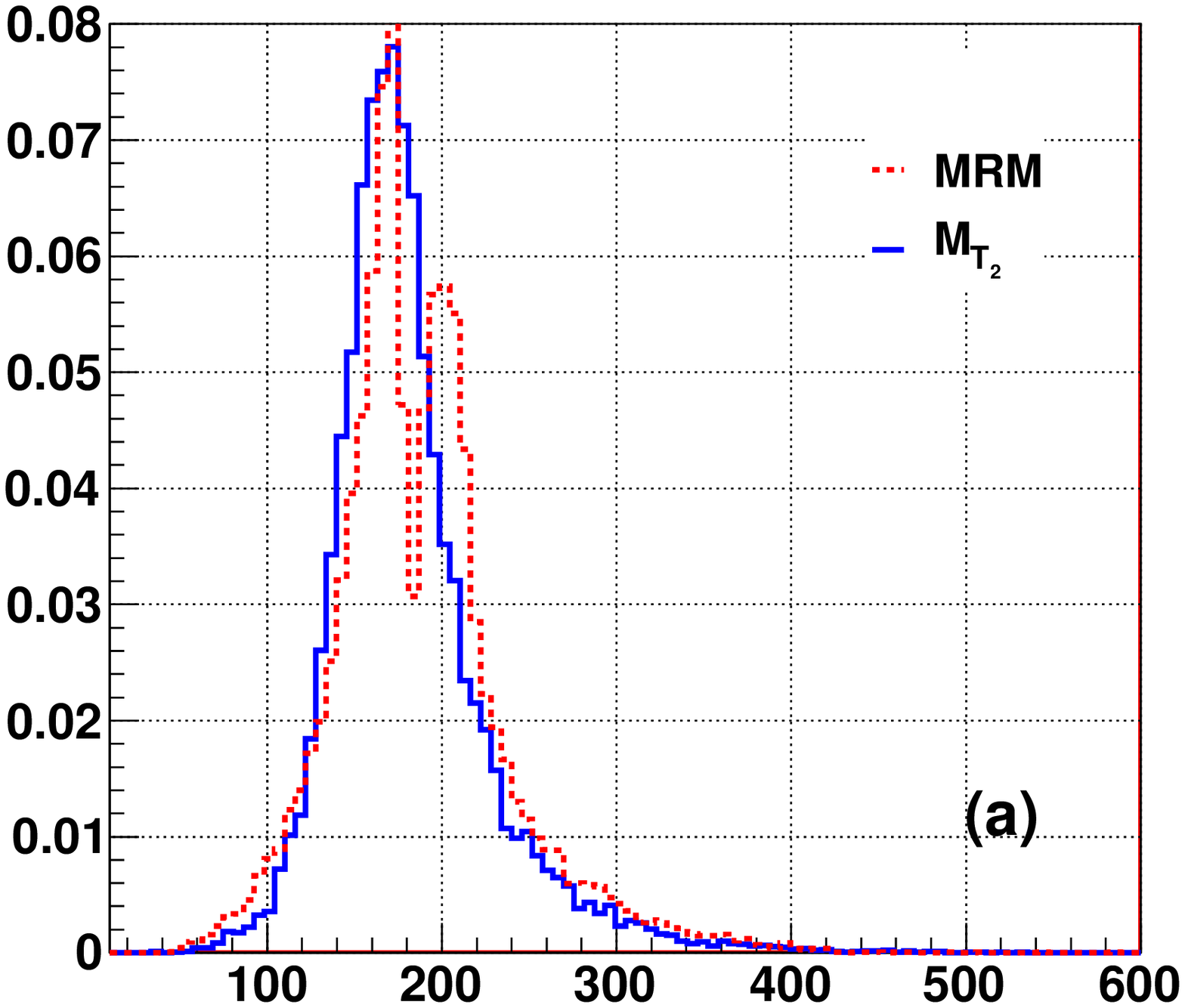}
\includegraphics[angle=0, width=.4\textwidth]{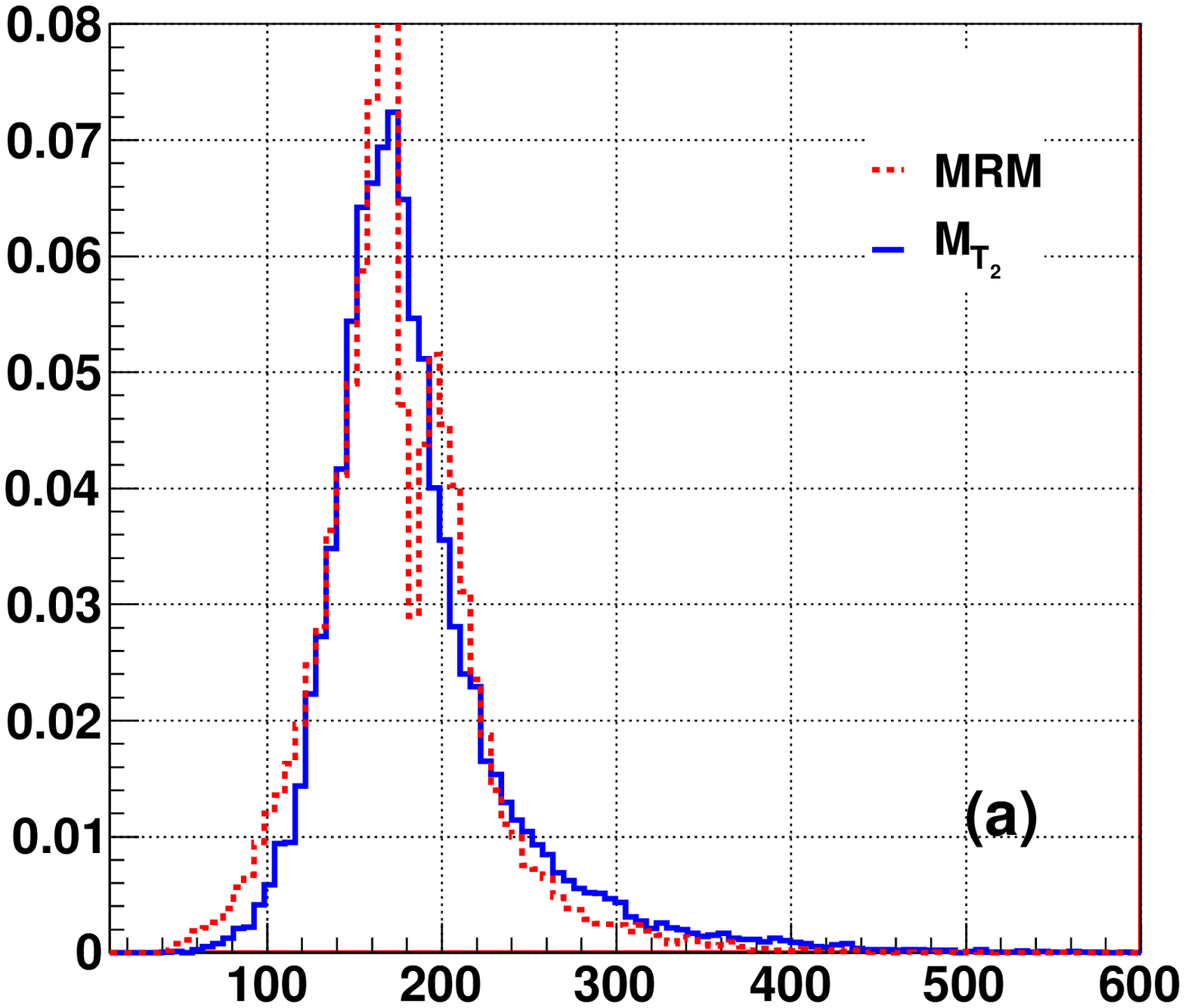}
\includegraphics[angle=0, width=.4\textwidth]{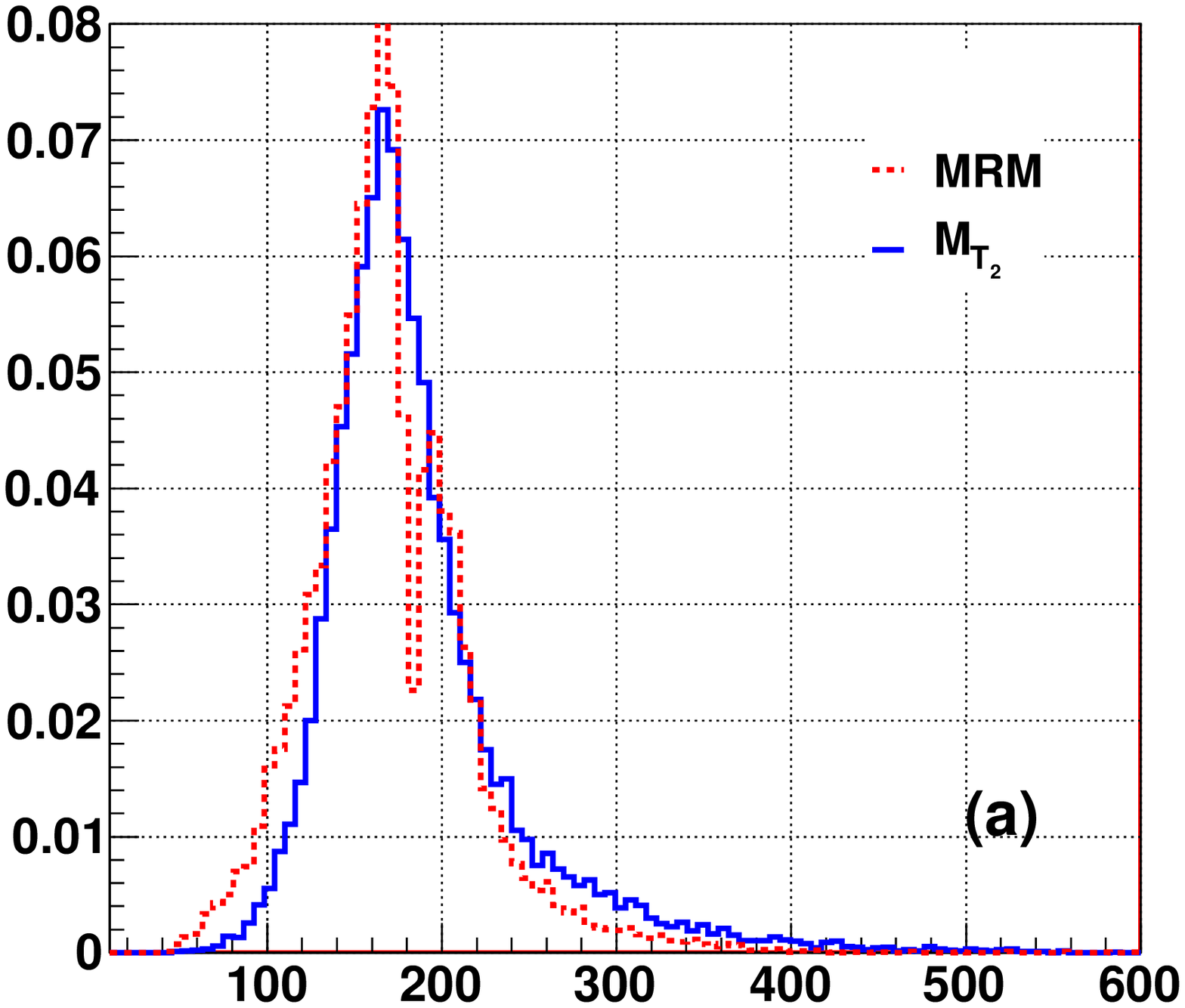}
}
\caption{\small\sf Reconstructed top mass using the variable $m_{T_2}$ 
for the event samples with $m_{Z'} = 0.5, 1$ and $1.5$ TeV in Figures 
(a), (b) and (c) respectively. $\sqrt{S} = 7$ TeV is assumed here.}
\label{fig:lhc7mt}
\end{figure}

\begin{figure}
\centerline{
\includegraphics[angle=0, width=.4\textwidth]{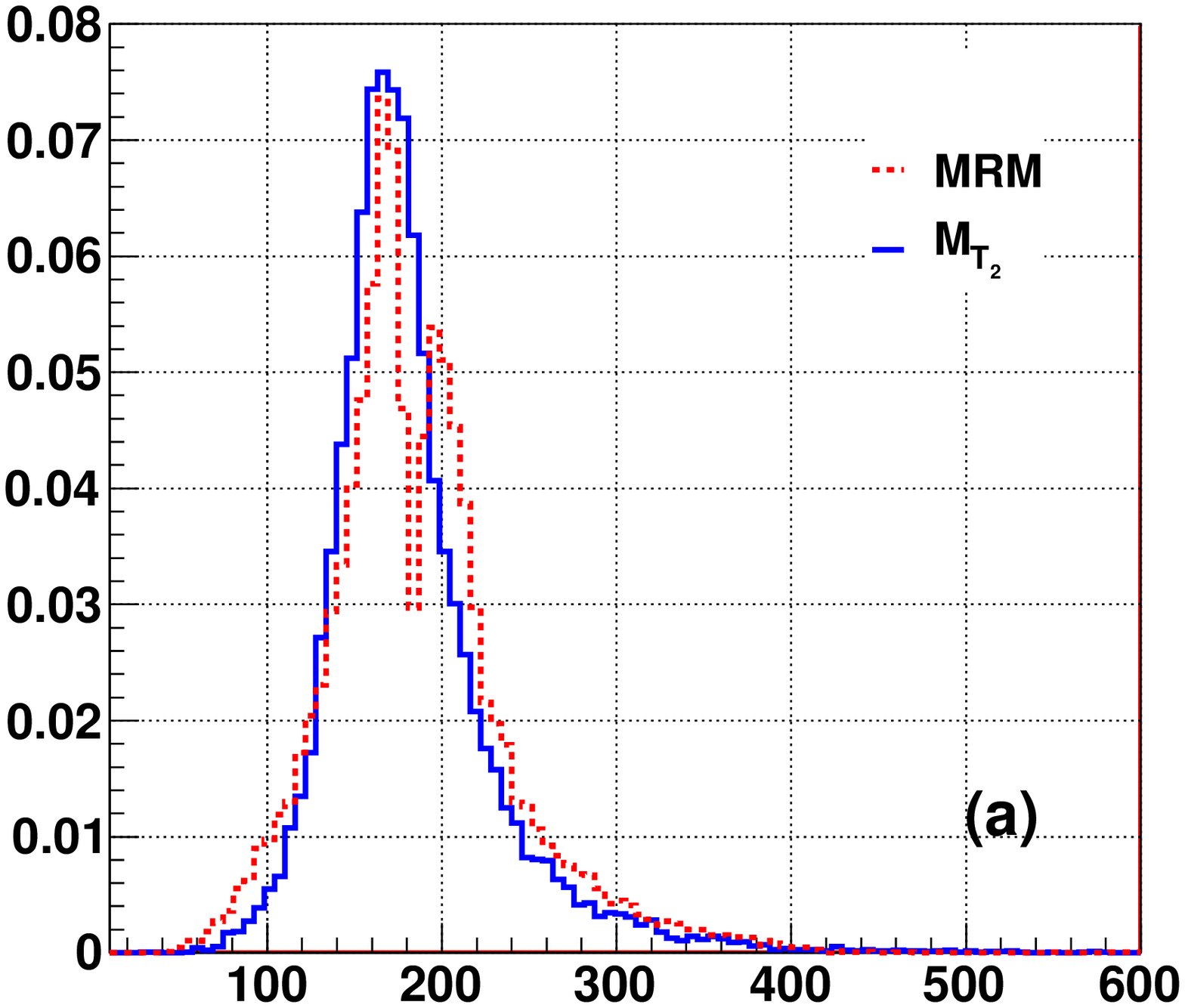}
\includegraphics[angle=0, width=.4\textwidth]{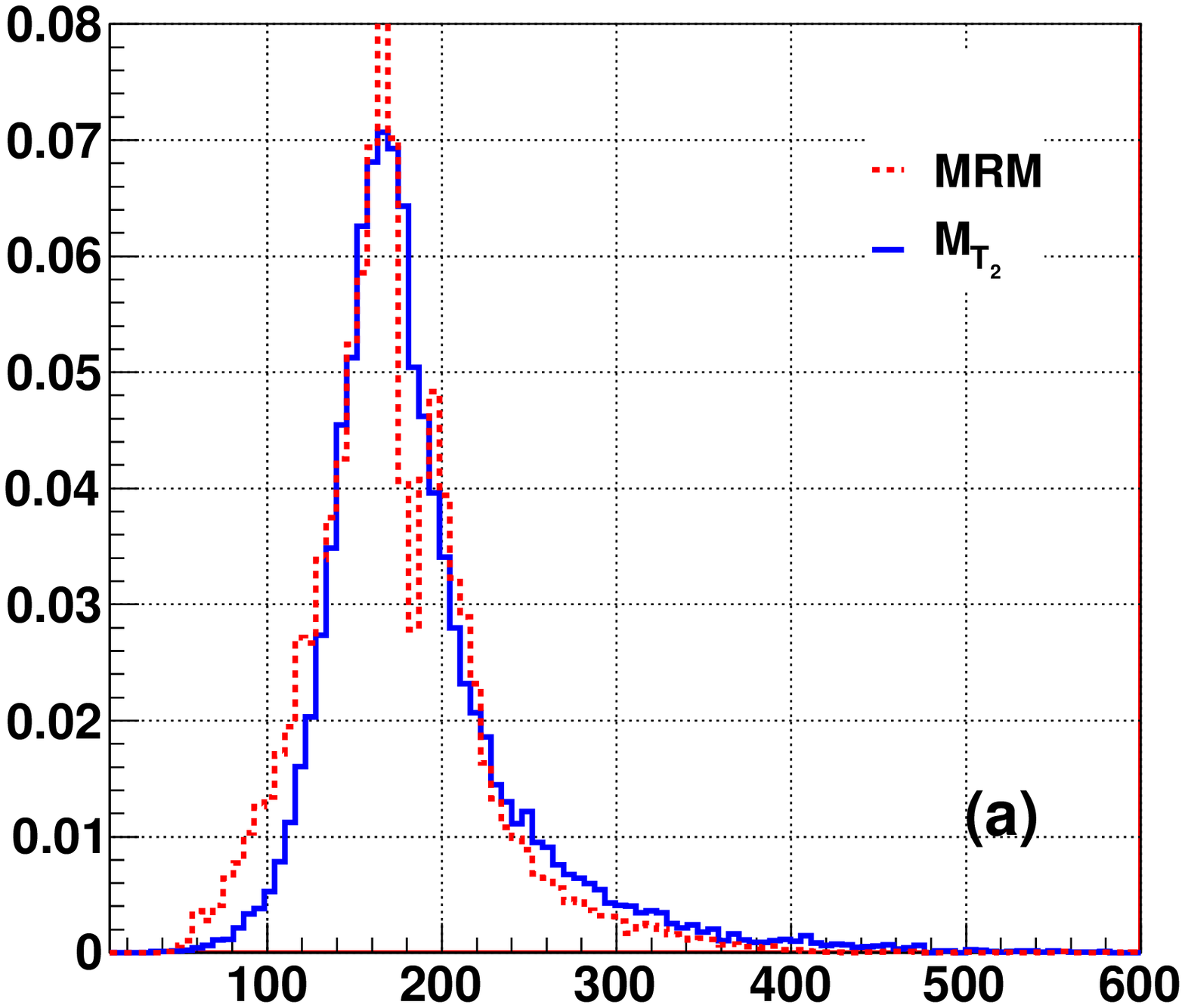}
\includegraphics[angle=0, width=.4\textwidth]{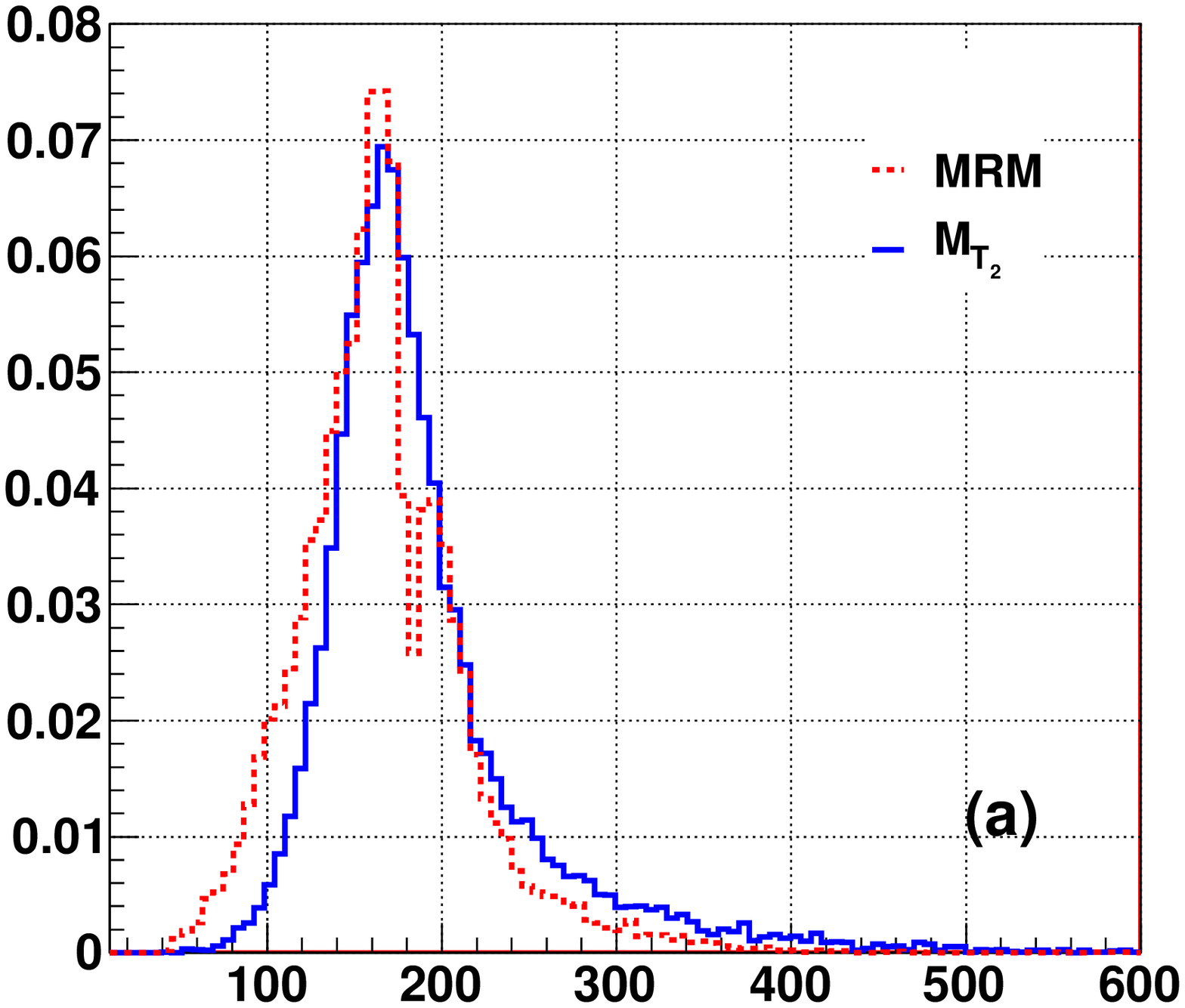}
}
\caption{\small\sf Reconstructed top mass using the variable $m_{T_2}$ 
for the event samples with $m_{Z'} = 0.5, 1$ and $1.5$ TeV in Figures 
(a), (b) and (c) respectively. $\sqrt{S} = 10$ TeV is assumed here.}
\label{fig:lhc10mt}
\end{figure}

\begin{figure}
\centerline{
\includegraphics[angle=0, width=.4\textwidth]{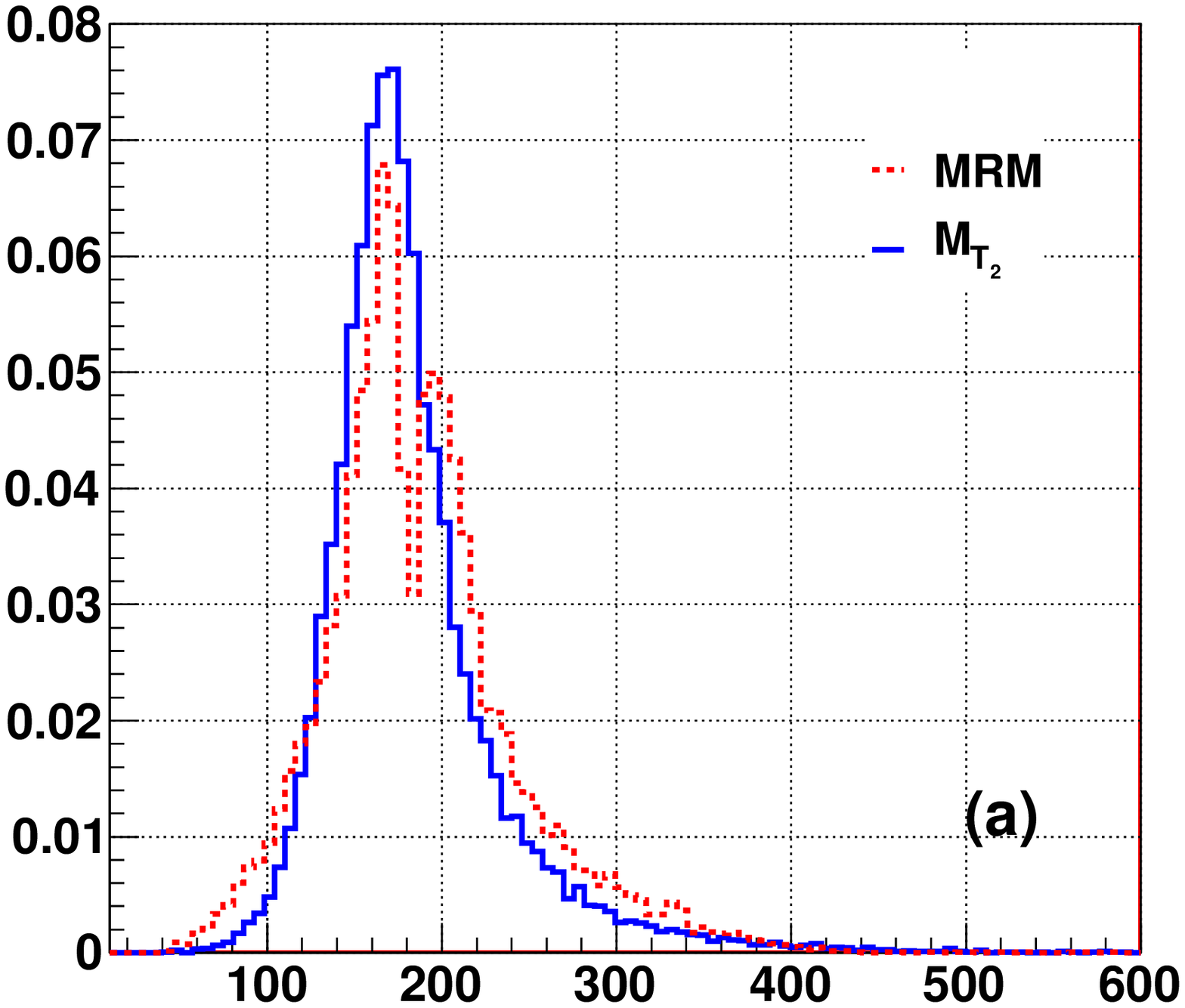}
\includegraphics[angle=0, width=.4\textwidth]{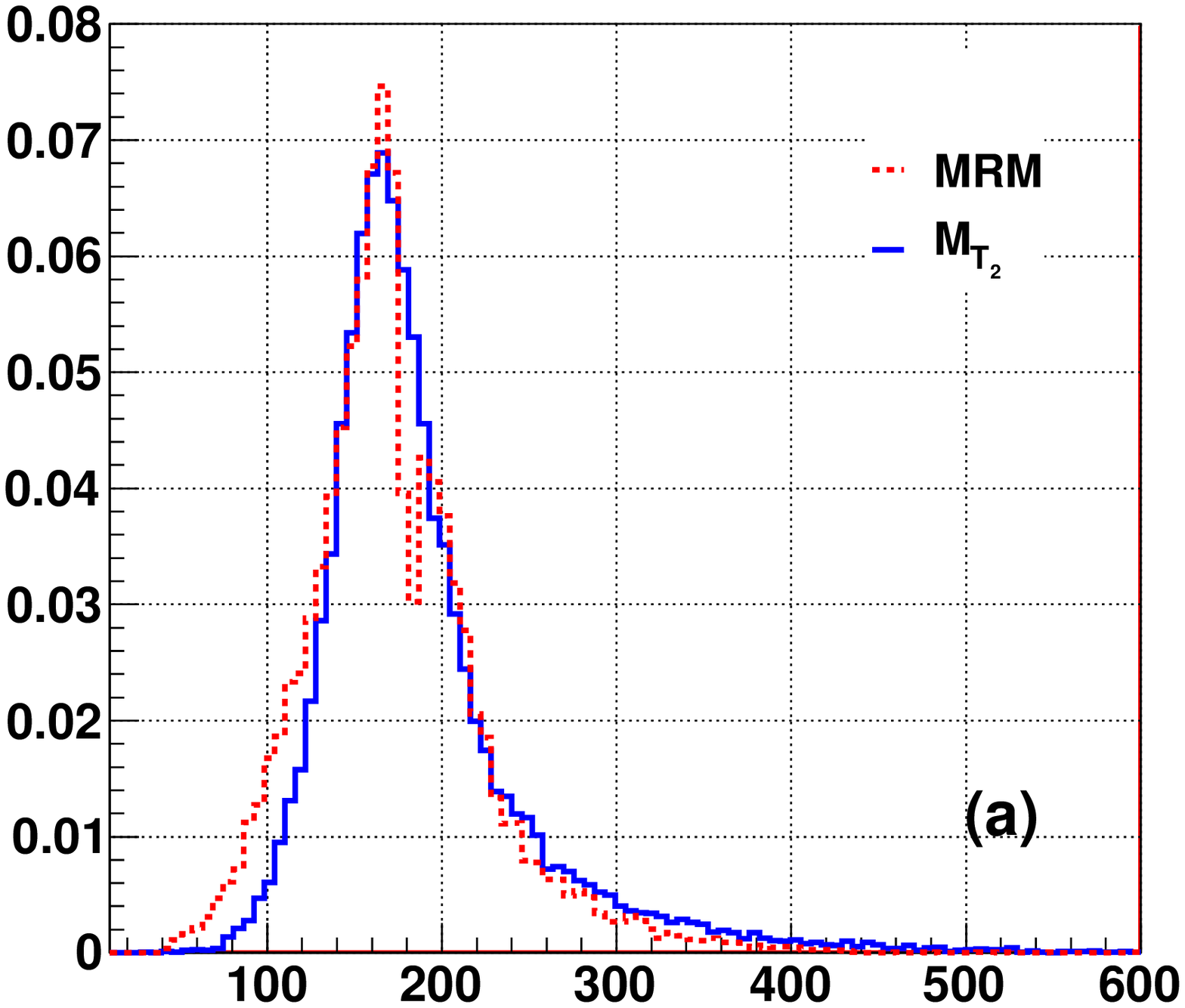}
\includegraphics[angle=0, width=.4\textwidth]{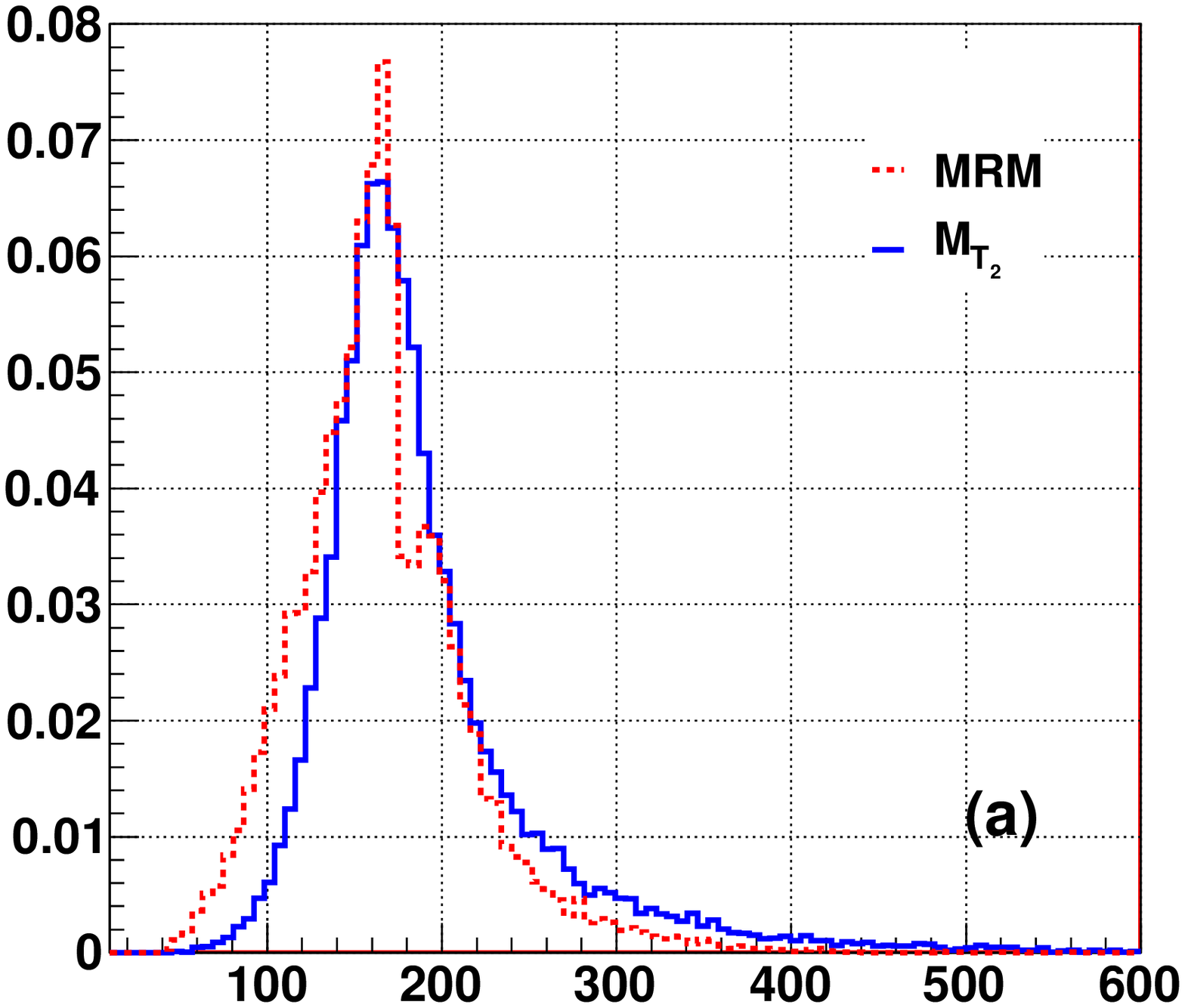}
}
\caption{\small\sf Reconstructed top mass using the variable $m_{T_2}$ 
for the event samples with $m_{Z'} = 0.5, 1$ and $1.5$ TeV in Figures 
(a), (b) and (c) respectively. $\sqrt{S} = 14$ TeV is assumed here.}
\label{fig:lhc14mt}
\end{figure}

We present reconstructed top mass in 
Figs.~\ref{fig:lhc7mt}-\ref{fig:lhc14mt} using both the methods. It is 
clear that the $M_{T_2}$ method does a little better job which is due to 
the fact it requires lesser information than the mass relation method. 
But at practical level, these hardly differ for the process under 
consideration, though, of course, the former is very helpful specially 
in longer decay chains such as in supersymmetry, universal extra 
dimensions and little Higgs models.

\subsection{$Z^{\prime}$ Spin Measurement}

\begin{figure}
\centerline{
\includegraphics[angle=0, width=.4\textwidth]{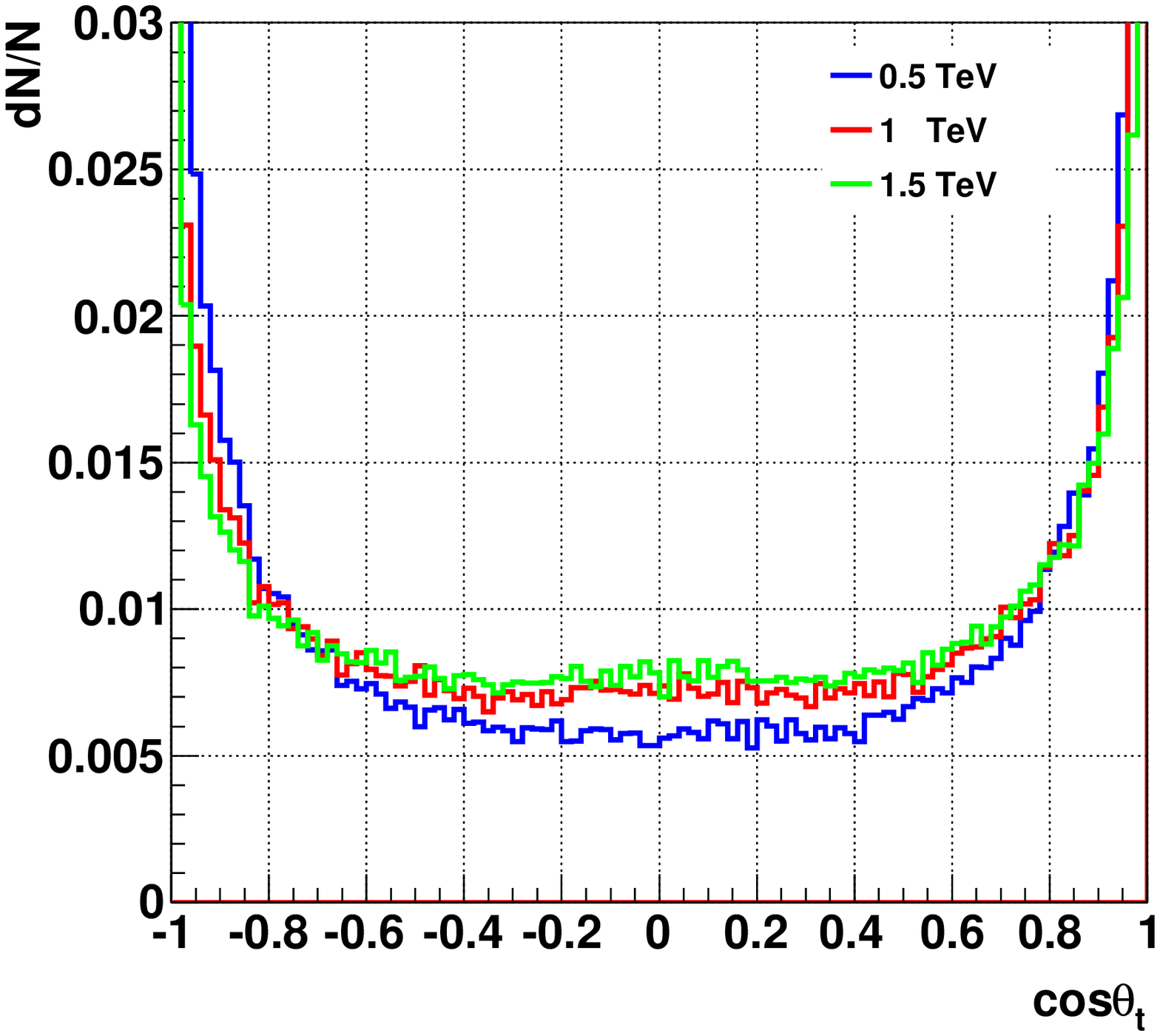}
\includegraphics[angle=0, width=.4\textwidth]{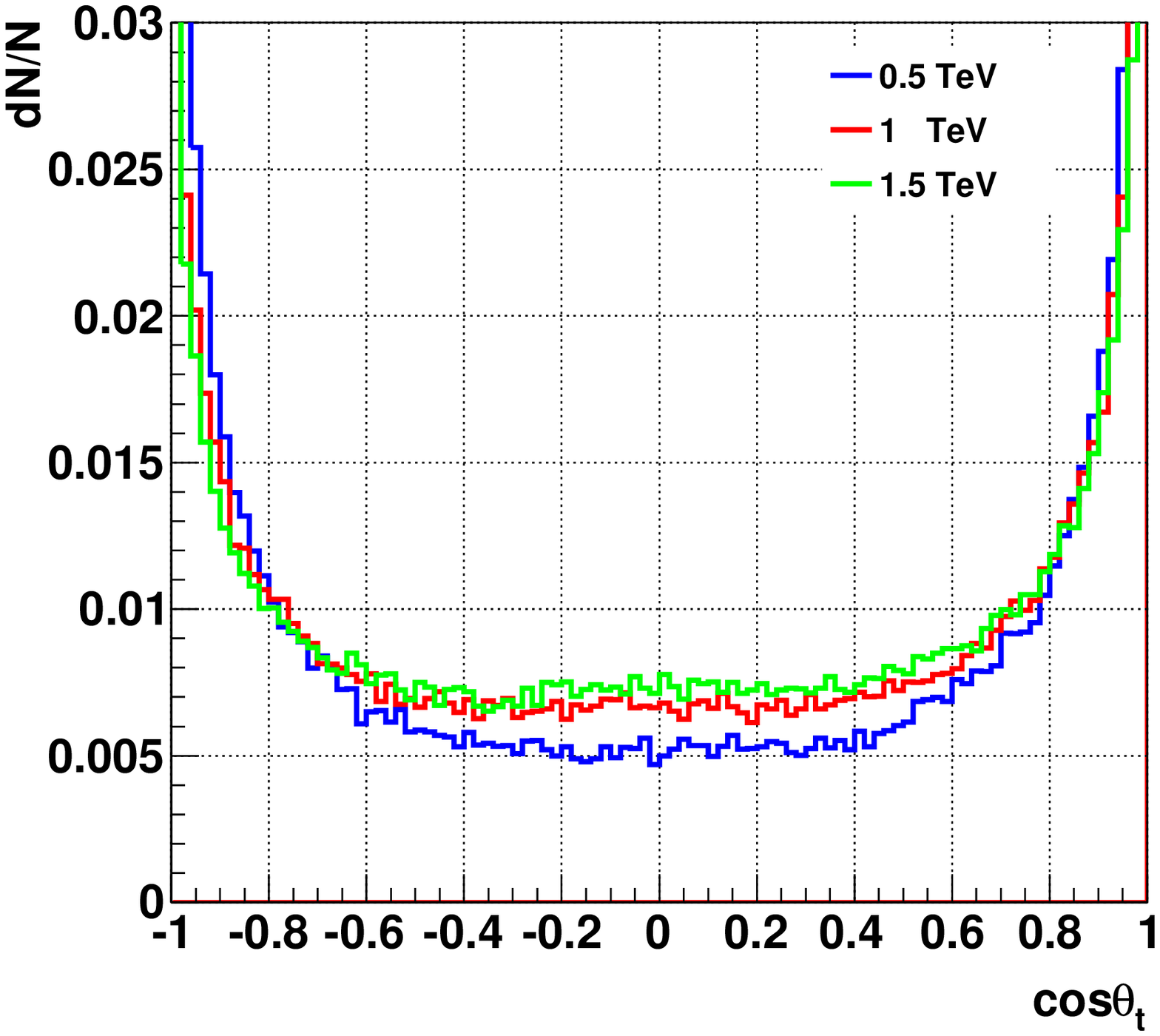}
\includegraphics[angle=0, width=.4\textwidth]{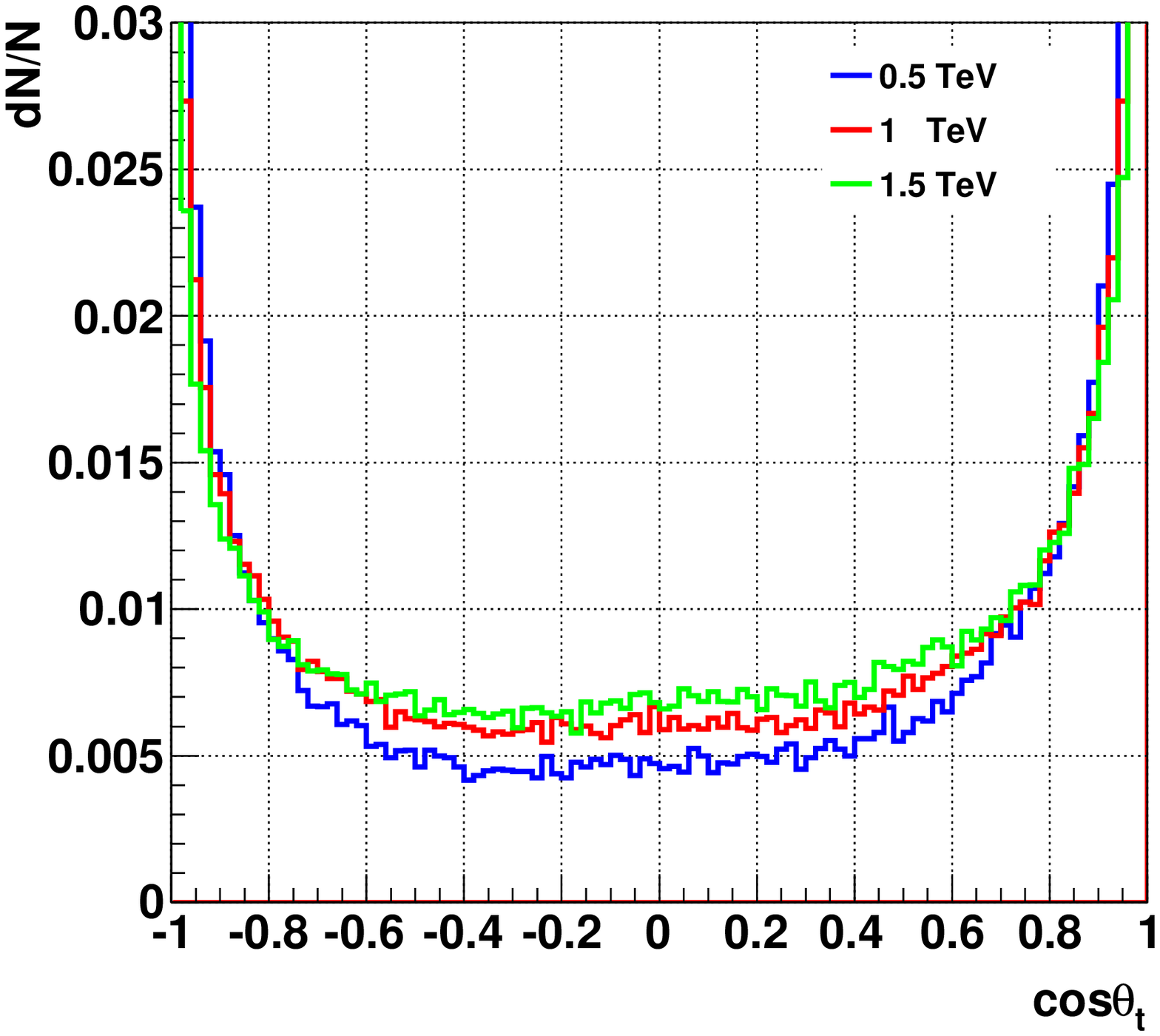}
}
\caption{\small\sf Angular distribution of the top in $tt$ CM frame with 
$\sqrt{S} = 7, 10$ and $14$ TeV.}
\label{fig:lhcspinz}
\end{figure}

With the reconstructed momenta as obtained in the previous section, and 
hence the $\sqrt{\hat{s}}$, we can fully reconstruct the partonic 
process just like a $e^+e^-$ collider. To gain more insight of the 
process, we investigate the angular distribution of the top in the 
parton CM frame. Results are presented in Figs~\ref{fig:lhcspinz}. We 
note that dip gets smaller with the rising $\zp$ mass, which will hint 
towards $\zp$ mass in addition to confirming vector nature of the 
exchange particle.  As has been established in 
Ref.~\cite{Godbole:2010kr} that using polarized tops, a right handed 
$\zp$ can be distinguished from the left-handed for at least up to $\zp$ 
mass of 750 GeV or so.

\subsection{LHC sensitivities to coupling}

\begin{figure}
\centerline{
\includegraphics[angle=-90, width=1.0\textwidth]{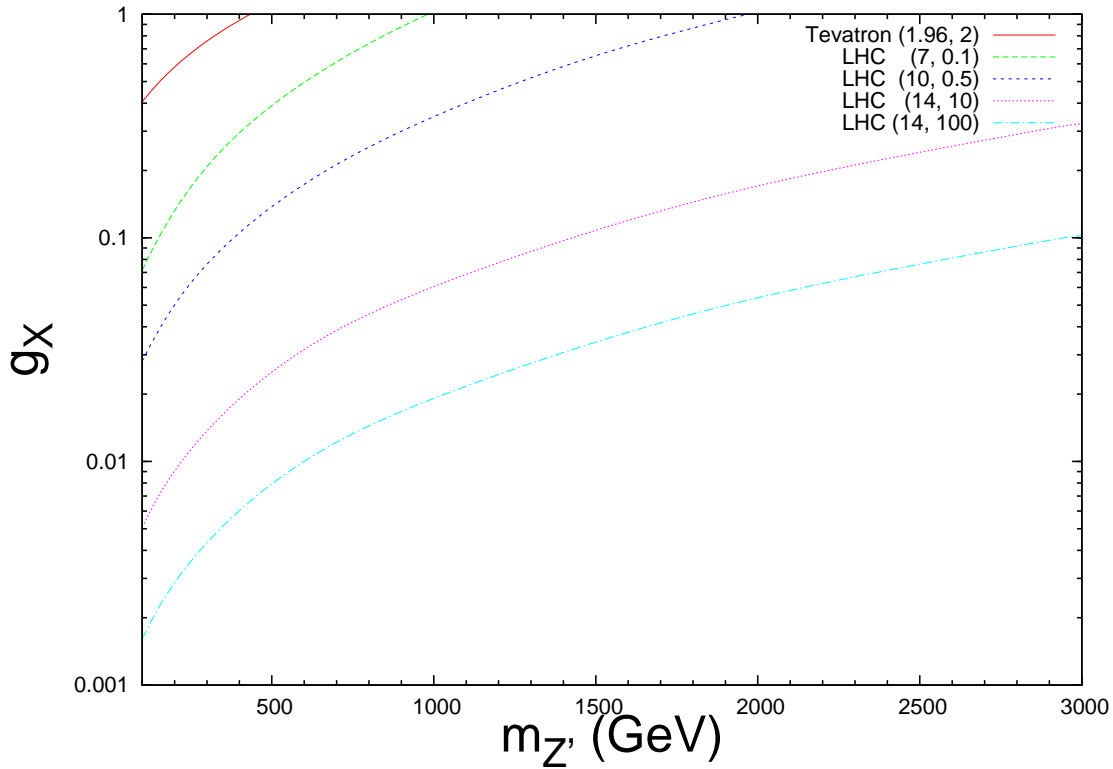}
}
\caption{\small\sf Allowed $g_{_X}$ - $m_{Z'}$ paramter space at the LHC 
and Tevatron in same-sign dilepton signal. In brackets are the 
Centre-of-mass energies (TeV) and integrated luminosities ($fb^{-1}$) 
respectively. Region right to each curve corresponds to at least 5 
events for the given luminosities.}
\label{fig:exclusion}
\end{figure}

In order to estimates the LHC sensitivities we assumed that at least 5 
same sign dilepton events are observed corresponding to each LHC energy 
we discussed. In Fig.~\ref{fig:exclusion} we plot these for a $\zp$ mass 
of up to 3 TeV. In the figure, region right to each of the curve is 
expected to be observed at the LHC besides what is already excluded at 
the Tevatron~\cite{Aaltonen:2008hx} as shown in the same Figure. As an 
example: For one year of LHC run (or equivalently saying, with 10 
fb$^{-1}$ data) with $\sqrt{S} = 14$ TeV, the lowest $g_{_X}$ that can 
be accessed, is $\sim 5 \times 10^{-3}$ which will further improve by a 
factor $1/\sqrt{\int {\mathcal L dt }}$ as more and more data is 
collected. One more thing to note that once we include the NNLO-NLL QCD 
K-factor as given in~\cite{Kidonakis:2003sc}, the lower allowed values 
of the coupling $g_{_X}$ will go down by a factor of $1./\sqrt{1.2} \sim 
1.1$, for a given $m_{Z^{\prime}}$.

\section{Results and Discussion}

We studied same a $\zp$ model that couples to the top quark with flavor 
off-diagonal coupling in the context of same sign dilepton signatures at 
the LHC with different LHC energies. We also estimated ratios of -ve 
signed dilepton with +ve signed dileptons as found that these can serve 
as an important tool in accessing the $\zp$ mass. We also reconstructed 
top mass using two techniques namely, through the on-shell mass relation 
method and the {\em $M_{T_2}$-Assisted On-Shell (MAOS) Momentum} 
technique and have shown that angular distributions of tops can be 
helpful in finding the nature of the exchanged $\zp$.

In the previous subsection, we also estimated LHC sensitivities to 
coupling with different LHC energies for $m_{z^\prime}$ up to 3 TeV. Our 
results shows that for an integrated luminosity of about $\int {\mathcal 
L dt } = 10 fb^{-1}$, the lowest coupling that can be reached using this 
signal is $\sim 5 \times 10^{-3}$ or so.

Further studies to measure top polarization~\cite{toppol} and various 
angular correlations~\cite{angcor} between the final states can be of 
paramount importance to understand nature of such $\zp$ and its coupling 
to the quarks.

\begin{acknowledgments}

SKG thanks German Valencia and David Atwood for useful discussion and com- 
ments, Nils Krumnack and Aranzazu Ruiz Martinez for some help with the ROOT 
package~\cite{Brun:1997pa}. The work was supported in part by DOE under contract number 
DE-FG02-01ER41155.
\end{acknowledgments} 
 

\end{document}